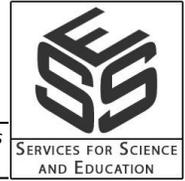

# A Survey of Security Threats and Trust Management in Vehicular Ad Hoc Networks

**Rezvi Shahariar**
Institute of Information Technology,
University of Dhaka, Dhaka, Bangladesh

**Chris Phillips**
School of Electronic Engineering and Computer Science,
Queen Mary, University of London, London, UK

## ABSTRACT

This paper presents a survey of state-of-the-art trust models for Vehicular Ad Hoc Networks (VANETs). Trust management plays an essential role in isolating malicious insider attacks in VANETs which traditional security approaches fail to thwart. To this end, many trust models are presented; some of them only address trust management, while others address security and privacy aspects besides trust management. This paper first reviews, classifies, and summarizes state-of-the-art trust models, and then compares their achievements. From this literature survey, our reader will easily identify two broad classes of trust models that exist in literature, differing primarily in their evaluation point. For example, most trust models follow receiver-side trust evaluation and to the best of our knowledge, there is only one trust model for VANETs which evaluates trust at the sender-side unless a dispute arises. In the presence of a dispute, a Roadside Unit (RSU) rules on the validity of an event. In receiver-side trust models, each receiver becomes busy while computing the trust of a sender and its messages upon the messages' arrival. Conversely, in the sender-side class, receivers are free from any kind of computation as the trust is verified at the time the message is announced. Also, vehicles can quickly act on the information, such as taking a detour to an alternate route, as it supports fast decision-making. We provide a comparison between these two evaluation techniques using a sequence diagram. We then conclude the survey by suggesting future work for sender-side evaluation of trust in VANETs. Additionally, the challenges (real-time constraints and efficiency) are emphasized whilst considering the deployment of a trust model in VANETs.

## INTRODUCTION

Vehicular Ad Hoc Networks (VANETs) allow emergency event dissemination and periodic announcements among road users. Vehicular networks promote the implementation of an Intelligent Transportation System (ITS) to reduce traffic congestion and improve the travelling experience for road users. Vehicles/drivers can announce false messages besides trustworthy messages. Trust management distinguishes the trustworthy messages from false messages. Hence, trust management for VANETs remains an important research domain. Many trust models are presented by numerous researchers for vehicular networks. Each of these trust models has a different trust evaluation strategy relying on a different set of trust metrics. Therefore, it is important to present the existing trust models in a condensed form. Many researchers review trust models in a differently. We also analyse existing trust models to highlight their strengths and weaknesses so that the readers can conveniently comprehend their characteristics. They can also learn about the different techniques which are employed in trust management for VANETs. From there, new researchers can consider new trust evaluation techniques in order to design an efficient trust model. Also, they may focus on different ways a particular technique is applied to develop a new model. For instance, reference [1] uses a Tamper Proof Device (TPD) to perform the trust computation. Conversely, reference [2] uses Tamper Proof Module (TPM) to manage the assigned credit score to determine the transmission cost and to generate signed messages.

Many surveys on trust management for VANETs are presented by different researchers. These surveys differ based on the coverage of approaches and the surveys are selective. Furthermore, some surveys are now outdated. We study existing literature, review many trust models, and compare recent trust models using a set of selected criteria where we demonstrate the achievements of many trust models in tabular form.

To ensure traffic safety, VANETs require trustworthy and accurate information sharing among the road users. If the sender is trusted, then it has a greater chance to announce a trustworthy message. A VANET should employ a malicious information detection method to identify untrue information disseminated over the network and to convey new messages to the users to disregard the false information as early as possible. Besides this, VANET must employ a communication scheme to alert vehicles in a timely manner for better traffic management. There can be a reward mechanism to encourage positive behaviour. Alternatively, punishment provision can discourage bad drivers and prevent further malicious activities from the dishonest vehicles. Receiver vehicles validate the messages and/or sources using a computation method to verify the accuracy of a message. In this way, vehicles determine whether a false message is sent across the network. This malicious information may influence unnecessary detouring of some vehicles and in some cases, they may reach a hazard zone forming a queue surrounding an event which results in severe traffic congestion. This frustrates the drivers who relied on the false information and took the earlier detour. A trust model can identify these false messages by evaluating the trust of the message and/or its sender which helps thwart the spread of false information and other attacks. Nearly all trust models evaluate messages after the arrival of messages. While verifying traffic events, some models collect recommendations or feedback data at RSUs or recipient vehicles. When an RSU collects data, it can disseminate a corrective message after the detection of the untrue information. If vehicles gather endorsements or indirect trust or opinion data, they individually determine the trustworthiness of an event. However, as RSUs have broader knowledge and are stand-alone in the event area, they can gather more information than the moving vehicles which results in more precise detection of untrue information than purely vehicular detection schemes.

There are many trust models which mainly focus on trust besides security and privacy issues for VANETs. Also, many surveys review well-established trust models, security, and privacy schemes. This provides an opportunity for a new survey to be presented to readers covering the new trust models along with the older ones which may help researchers to discover new knowledge after examining the existing research. As trust models deal with security threats and attacks in VANETs, many surveys begin by explaining known threats. After this, they summarize existing trust models and classify them considering a set of selected criteria, for example, scalability, distribution, robustness, and ability to thwart particular attacks. This is typically followed with suggested future research directions which helps researchers develop new trust models. For example, in [3], VANET elements and the form of communication are described first. Then existing trust models are reviewed. They focused only the papers published between 2017 to 2022. They discuss these models and then highlight the challenges for security and trust in VANETs. Furthermore, they discuss the future direction of trust management research. In [4], the authors compare and review trust models from 2014 to 2019. Additionally, they present the weaknesses of some trust models. They also suggest some challenges that need to be tackled to achieve trustworthy communication in VANETs.

Trust management is crucial for vehicular communication. As both untrue events and trustworthy events can be announced. So, it is necessary to distinguish between trustworthy and untrusted messages. To this end, security and trust models are employed to confirm the trustworthiness of messages. However, when an authorised user sends untrue information, security approaches fail. Security can address authentication, authorization, availability, data tampering, integrity, and nonrepudiation. However, a trust model is used to isolate misbehaviour from authorised users which enriches the security enforcement by allowing the available services to be restricted to trusted users. Untrusted sources are excluded from service interaction. In this way, VANETs remain secure from malicious activities as these are thwarted by trust and security model. Additionally, the sender of false information needs to be penalised. A trust model evaluates the disseminated information to reward positive behaviour or to punish negative behaviour. In this way a model helps encourage the trust of a sender. Provided a sender maintains a good trust score, it is expected that it will send trusted messages in the near future. Nevertheless, a trusted sender can lie in any situation which also needs to be detected for the assurance of reliable information dissemination. Some vehicles build trust, and others lose it over time depending upon their announcement characteristics. If a trust model is implemented in both vehicles and RSUs, the trust manager can monitor the trust of every entity.

**Structure of this Survey**
In this paper, a survey of existing trust models is presented which starts listing many known threats for VANETs as trust management is used to thwart internal attacks. We review trust models presented from 2005 to 2024. The selection criteria for this survey are the wide acceptance of the paper (the number of citations a paper receives), the publication year, and the technological variation relative to other trust models. After this, we review each trust model, including its strengths and weaknesses. Next, trust models are classified considering many features, for example, data collection method, evaluation process, and the technology used by the respective trust model (i.e. blockchain, machine learning and so forth).

Furthermore, some features are summarised in tables using a selected set of metrics, for instance, whether the trust evaluation scheme needs to collect any kind of recommendation/feedback data from the neighbouring vehicles/RSUs. As this metric affects the communication overhead and can delay the decision-making process. Also, the summarization demonstrates the different metrics a trust model requires to evaluate the trust. Furthermore, we include the capability of a trust model in terms of security threat handling. The paper also summarizes the analysis each trust model compared to a baseline scheme. Finaly, it examines the simulated scenario(s) considered by trust model and which simulators are used for the evaluation. Therefore, we believe readers will obtain a good understanding of existing trust models from the comparative tables. Here, we apply filtering of the trust models, discarding trust models which employ similar mechanisms to keep the table size manageable.

**VANET Components**
In VANETs, Roadside Units (RSUs), vehicles fitted with an Onboard Unit (OBU), and optional Central/Trust Authority (CA/TA) are connected in a network. Vehicles encounter each other randomly. Vehicles move quickly (1km/s to 120Km/s) so their topology changes very dynamically. Their communication range is within 0~1000m distance, and the topology is more dynamic than other forms of ad hoc network. Vehicles and RSUs typically communicate using the IEEE 802.11p-based Dedicated Short-Range Communication (DSRC) protocol. A dedicated bandwidth of 5.9 GHz is reserved for the communication between elements in VANETs [5]. Every type of vehicle including regular vehicles, official vehicles (ambulance, police, firefighting vehicles) and public service vehicles (buses, and licensed taxis) can be part of a VANET. They expect to obtain timely traffic updates from the nearby RSUs. Fig. 1 illustrates a standard VANET where all the major components are shown.

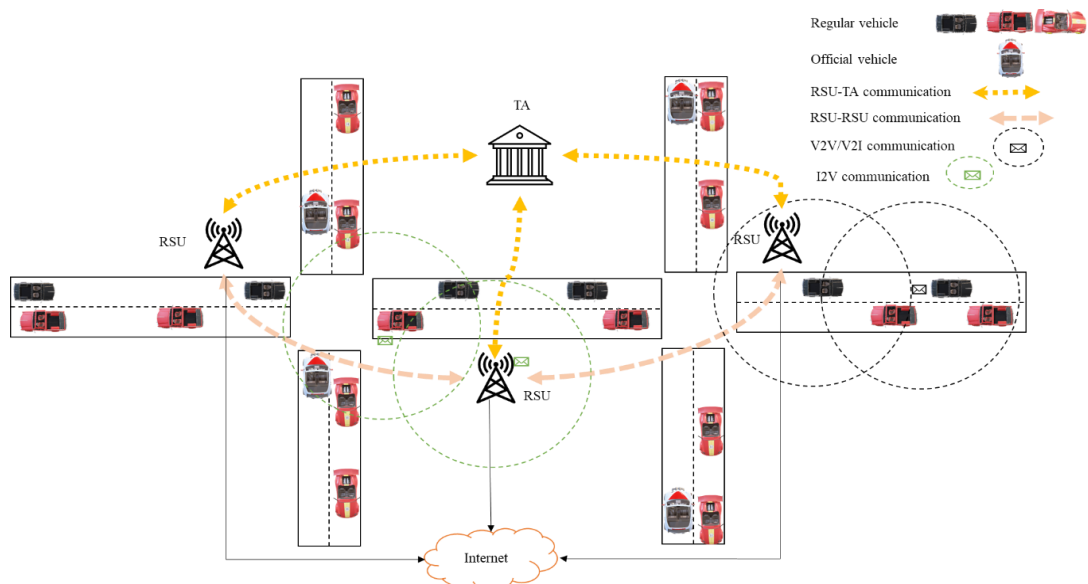

**Fig 1: Typical VANET Scenario with its Major elements [6]**

Regular vehicles are the key users of a VANET. They can announce periodic beacons and emergency traffic events. Vehicles announce traffic incidents only when they notice any incident on the road. A receiver vehicle is the one which receives event information from another vehicle. Upon reception, they may act on the event or relay it in the vicinity so that adjacent vehicles are updated concerning the traffic event. An intermediate vehicle that relays an event is called a forwarder or relayer. In this manner, one vehicle can assist neighbouring vehicles from being stuck in congestion or an unwanted situation on the road. Every vehicle is pre-equipped with an On Board Unit (OBU) which includes a transceiver to communicate with other vehicles and RSUs [7]. Furthermore, the OBU may have a Global Positioning System (GPS) sensor and an Event Data Recorder (EDR) [7]. It is also usual to install a Tamper-Proof Device (TPD) to store records or to perform some computation inside the vehicles [8, 9]. Many classes of official vehicles may be present on the road. Among them, ambulance, police, and firefighting vehicles are the most common. They visit an event place to assist when they are instructed. When the incident is resolved, they usually send a message stating the resolution so other vehicles can use that road again. Incident information from them can be deemed authentic. RSUs are positioned alongside the road to broadcast timely traffic updates to vehicles. They can communicate between themselves either using wired or wireless communication. Moreover, they are linked to the Central/Trust Authority (CA/TA) through a dedicated wired or wireless Internet connection. RSUs update traffic incident information to the CA/TA. Road users obtain periodic traffic broadcasts as well as emergency incidents from RSUs [7, 10]. Additionally, RSUs consider messages from official vehicles as "high priority" when they are attending to specific emergencies. The Trust/Central Authority (TA/CA) is the ultimate authority in a VANET. The TA registers RSUs/vehicles, authenticates vehicles, and blacklists malicious vehicles when the extent of their maliciousness goes beyond a threshold of poor behaviour. It is obligatory to place the TA in a highly secure environment. Furthermore, the TA must be provided with adequate computing resources to fulfil the processing demands from other entities.

A vehicle broadcasts a message which is relayed by adjacent vehicles to reach non-neighbouring vehicles. Direct communication takes place when vehicles send periodic events, for example, beacon messages containing the status (position, direction, velocity) of the vehicle to other vehicles. Multiple-hop communication is required whenever vehicles disseminate emergency traffic events to others. When there is no neighbouring vehicle, announcement messages are dropped. Broadcasting of event messages at the right time enables detouring to alternate paths and avoiding traffic congestion or other undesirable phenomena. In VANETs, when a vehicle notices an event, it broadcasts a message, and other vehicles relay the same information until a specific condition is met. When this event message reaches an RSU, it also periodically broadcasts the same until the event is resolved. In some severe cases, RSUs share traffic events with neighbouring RSUs so that more vehicles will not enter the problematic area from adjacent regions. In this way, severe traffic chaos can be avoided. Also, vehicles and RSUs periodically send beacon messages using single-hop communication which contain speed, direction, location, and acceleration to update their status information [10]. In contrast, traffic events are announced using multi-hop communication. Multi-hop communication facilitates communication among vehicles outside the transmission range of the originator. In this situation, intermediate vehicles simply forward the messages [11]. Additionally, group communication is common inside a cluster in a VANET.

A CA/TA holds records about events, registered and access-blocked vehicles/drivers. Besides this, the network allows the dissemination of traffic situations neighbouring vehicles and RSUs. These cover dissemination of all kinds of traffic incidents including but not limited to accidents, traffic jams/congestion, diversions, severe weather, and poor road conditions. This service warns other vehicles not to use a particular road or section of a road to avoid possible congestion. When messages arrive at receiver vehicles, they need to determine the validity of the message. Hence, the evaluation of the message is important to verify the source and/or message. A trust management scheme does this. Thus, the selection of a trust model is vital as it relates to the operating efficiency of the VANET. In [1], it is identified that two criteria are most important to achieve efficiency which are the communication overhead and driver decision time/response time. It is also shown in [1], that these two metrics are lowered with sender-side trust evaluation than for receiver-side evaluation of trust since in receiver-side systems all receivers independently need to find or communicate to a central server or each other to evaluate the trust of the message of the sender.

**Security Requirements for VANETs**
The security requirements for VANETs are availability, authentication, access control, privacy and confidentiality, integrity, non-repudiation, real-time constraints, as well as efficiency [12]. An attack is a threat to a system that facilitates an attacker to access, alter, add, delete, or disclose information without obtaining permission from the authority [13]. Attacks impede the normal operation of VANETs. An attack can be either active or passive. An active attacker alters the information from a target entity or changes the ongoing route to the target [14]. In contrast, a passive attacker only observes the network activities to discover confidential information to find vulnerabilities [14]. Also, attackers can initiate attacks from both inside and outside. An attacker is external when he/she is not authorized within the VANET. An outsider can listen in user communication to launch Distributed Denial of Service (DDoS), black hole, or false information injection attacks in VANETs [15]. An insider attacker is an authorized user of a network who can launch all possible forms of attack as he/she has gained wider knowledge about the network. Insider attacks cannot be thwarted using only a cryptographic scheme which can foil outsider attacks [16]. Hence, trust management is used as a second line defence to limit internal attacks. Additionally, trust management can classify honest and dishonest vehicles as well as revoke messages from malicious vehicles [17]. Furthermore, a trust scheme can achieve access control, mischievous vehicle isolation and allocate punishment for mischievous activities [18]. Despite this, trust models cannot protect VANETs completely [19] as it is not assured that a trustworthy vehicle will always stay trustworthy and announce trustworthy events. Hence, an evaluation model using predetermined metrics can determine the reliability of sent information, and reward or punish a sender vehicle appropriately [18]. By evaluating announcements, trust management can play a major role in identifying malicious vehicles and protecting the VANET from mischievous actions. Before reviewing and classifying trust models, it is worth considering the types of attacks that can be launched in VANETs. These attacks are grouped next based on the security threat which arises from a specific attack.

The term availability signifies that a system is functioning though a part of the system may be malfunctioning or faulty [20]. Channel availability is always required to achieve communication in VANETs otherwise, messages cannot be relayed between entities. Thus, availability must be satisfied for each element in a VANET. In Table 1, the third column includes a list of known attacks which may affect availability. Authentication is the verification of a user while attempting to obtain access to the system resources to which they are entitled. Authentication is the first level of defence against any kind of intrusion. For VANETs, three things are typically authenticated for VANETs which are: ID authentication (license plate, chassis number), property authentication (sender is a car or RSU), and location authentication [20]. Attacks that interfere with authentication are shown in the fourth column of Table 1. Privacy is one of the greatest challenges in a VANET. Despite anonymous communication the tracking of a real user is still possible [12]. Drivers wish to keep their private information secret [21, 22]. Privacy protection means the protection of location and identity information against profiling. Confidentiality ensures only the intended receiver can have access to the desired information. Thus, it is essential to allow anonymous communication besides trustworthy announcements in VANETs [23]. The objective is to develop a system where sensitive information cannot be accessed unauthorizedly [12]. Attacks against privacy and confidentiality are summarised in the fifth column of Table 1. Integrity confirms that the data in transit is not modified by any means and the receiver receives whatever the sender transmits. Integrity protects data from unauthorized creation, dismantling, and modification [12]. Integrity confirms an attacker cannot modify the meaning of a message so that the message remains reliable [24]. To confirm integrity, a digital signature is first created, then it is attached to the message by the sender. Once the message is received by receivers, they verify the digital signature prior to accepting the message. Integrity attacks are listed in the sixth column of Table 1. Nonrepudiation validates that either party involved in the communication cannot decline their involvement [12]. For example, sender nonrepudiation ensures that a sender cannot later refuse the fact that it sent a message. Threats to nonrepudiation are shown on the last column.

**Table 1: Security Threats for VANETs**

| Attack Name | Attack Definition | Threats to Availability | Threats to Authentication | Threats to Privacy | Threats to Integrity | Threats to Nonrepudiation |
|---|---|---|---|---|---|---|
| Denial of Service (DoS) attack [10] | DoS inhibits rightful users from getting access to resources to which they are entitled. Both insider and external perpetrators can make the wireless medium busy and prevent access to networking resources granted to legitimate users. Additionally, Distributed Denial of Service (DDoS) is the launching | ✔ | | | | |

| | of an attack in a distributed manner to disrupt normal network operations. | | | | | |
|---|---|---|---|---|---|---|
| Jamming attack [10] | The attacker makes the medium busy by using the same frequency and a stronger signal than the data signal during communication. | ✔ | | | | |
| Malware attack [10] | This infects a vehicle or an RSU to behave abnormally. | ✔ | | | | |
| Broadcast tampering attack [10] | Malicious authorized users add incorrect data to change the meaning of a message. | ✔ | | | | |
| Greedy behaviour attack [10] | When attackers abuse the precious bandwidth is a greedy behaviour attack. | ✔ | | | | |
| Spamming attack [10] | An attacker sends spam to collide with genuine messages to confuse recipients. | ✔ | | | | |
| Black hole attack [10] | They deny forwarding packets or forwarding to another direction to disrupt the established connection. | ✔ | | | | |
| grey hole attack [10] | Initially, they behave honestly to build trust but afterwards drop packets to take advantage. | ✔ | | | | |
| Sybil attack [10] | Malicious users create multiple fabricated identities and disperse false messages using different identities. | | ✔ | | | |
| Node impersonation attack [12] | When messages are received from a real user, an attacker changes and relays them with a stolen identity for his/her benefit. | | ✔ | | | |
| Tunnelling attack [10] | An attacker creates a long-distance tunnel to send/receive with a targeted remote vehicle and considers it as its neighbour. | | ✔ | | | |
| GPS spoofing attack [10] | An attacker creates a false GPS signal to betray other entities. | | ✔ | | | |
| Key and/or certificate replication [7] | The attacker uses the key and/or certificate of another vehicle to approve itself. | | ✔ | | | |
| Eavesdropping [10] | Eavesdropping is listening to any confidential information. | | | ✔ | | |
| Man-in-the-middle attack [20] | An attacker listens to the communication between a sender and a receiver and modifies messages. The attacker bedazzles the sender and receiver to trust that they are communicating directly, unaware of the altered messages. | | | ✔ | | |
| Home attack [20] | In three ways an attacker can launch a home attack. First, an attacker takes control of an OBU of a vehicle and then inserts incorrect messages in the VANET. Second, an attacker takes control of the sensor of a vehicle to alter the activities of the sensor. Third, an attacker takes control of the Electronic Control Unit | | | ✔ | | |
| | (ECU) of a vehicle and changes the speed of the vehicle. | | | | | |
| Social attack [20] | It is the act of persuading a driver to broadcast improper messages so that other drivers get irritated, and their driving behaviour is disturbed. | | | ✔ | | |
| Identity disclosure attack [13] | An attacker sends a message to a target vehicle rightfully to obtain a response from the target vehicle and to track it. Then the attacker divulges the identity and the location information. | | | ✔ | | |
| Traffic analysis attack [10] | An attacker analyses transmitted messages to capture confidential information. | | | ✔ | | |
| Message suppression/ fabrication attack [10] | In this attack, relaying nodes alter messages towards the destination to delude the intended recipient. As a result, some vehicles may detour using a longer route or get stuck in a traffic jam. | | | | ✔ | |
| Message falsification attack [12] | This attack occurs when incorrect information is disseminated. For example, it might be sending "an incident on road" message when there is none to mislead vehicles. | | | | ✔ | |
| Masquerading attack [10] | An attacker uses misappropriated information to get access and then broadcasts a false message. | | | | ✔ | |
| Replay attack [10] | This attack occurs when an attacker reinjects earlier messages to bedazzle others. | | | | ✔ | |
| Hardware tampering [12] | This attack results from intervening with the hardware of an entity in the VANET. | | | | ✔ | |
| Repudiation attack [10] | In a repudiation attack, a sender/receiver denies sending or receiving a message. | | | | | ✔ |

## TRUST MANAGEMENT FOR ENRICHED VEHICULAR SECURITY

Trust is the degree of belief that can be placed on the message from another vehicle. In VANETs, trust is established between vehicles based on the messages they exchange with each other over time. Trust management limits the dissemination of false information in VANETs [25, 26]. Trust is a key element of security which is defined as "A system or component that behaves expectedly for a particular purpose" [27]. For vehicular networks, trust is established among vehicles based on their honest or dishonest behaviour towards others which are evaluated by receivers whenever a message is announced or relayed to neighbours. Receiver vehicles calculate the trust of a sender, and its messages based on some predefined metrics, for example, past interaction(s) with the sender, recommendations, and the reputation of the vehicle [19]. Many different trust metrics are considered for evaluation of trust of the sender and its messages which are functional trust, role-based trust, indirect/recommendation trust, direct trust, reputation, honesty factor, cooperability, and time and location closeness to an event and so on. In addition, trust evaluation is extremely difficult because many factors need to be considered in a short time [28]. In [18], the authors noted that the trustworthiness of transmitted information is typically periodically assessed using predefined metrics and computational procedures. Whether trust is computed at the receiver vehicle side or the RSU is a design feature of a trust model. If trust is computed at the RSU then the concerned vehicle collects updated trust from the RSU. Also, the metrics required for the computation of final trust and their selection need to be defined. These decisions need to be carried out carefully to make the trust scheme robust, scalable, or efficient in terms of both communication and computation. A vehicle which sends trustworthy messages most of the time builds trust to others and vice versa. Vehicles upon message reception decide an action either by detouring to another route to avoid the probable congestion or move towards the planned route. Thus, trust can regulate vehicle movement to an alternate route. Hence, a mischievous driver can send a false message to detour many vehicles to a particular route to initiate severe traffic conditions to achieve his/her goal. These malicious messages need prompt detection to help vehicles to move freely on roads. This also demands timely message delivery to others. It is not guaranteed that a trustworthy vehicle will always announce truthful events. Hence, an evaluation is necessary either to reward or to punish a given source vehicle. Vehicles which steadily preserve a good trust level can be believed trustworthy to others as their current trust level implies that they announced trustworthy messages in the past. Thus, by evaluating announcements, trust management can play a major role in protecting the VANET from mischievous actions and help to identify the malicious vehicles based on their actions.

A VANET disseminates critical messages (accidents, traffic jams etc) towards vehicles and other entities. Hence, VANETs require a secure, trusted environment for the broadcasting of exact, trustworthy, and genuine information. It is extremely challenging to uphold this requirement because of the large scale and open environment which is sensitive to various forms of attack. A cryptographic approach cannot safeguard the VANET wholly due to attacks from authorized users. To thwart attacks from authorised users, trust establishment has been embraced [19, 29]. Trust and reputation schemes enrich security by promoting good behaviour and penalizing malicious behaviour [30]. In the next subsections, trust models are briefly reviewed with their positive and negative aspects.

**State-of-the-Art Trust Model Classification**
Previously, most trust models have been presented for mobile ad hoc networks [31] and wireless sensor networks [32]. However, they cannot be immediately applied to a VANET due to its characteristics and requirements. Therefore, trust models for VANETs remains an active research topic although many trust models have already been proposed [19]. In this section, some well-known trust models are classified and reviewed. Trust approaches aim to classify honest and dishonest vehicles. Several schemes adopt access-blocking of malicious vehicles when their behaviour is proven extremely bad while others allow constrained announcements when the trust levels of vehicles are low. Trust approaches can be classified in several ways. For example, some approaches validate the trust of vehicles in a centralized server whereas others compute trust at the vehicles in a decentralized manner. Additionally, they can differ in their data collection mechanism or the technique they use for trust management, for example, machine learning, fuzzy logic, blockchain, or probabilistic and statistical mechanisms are already used in trust management. We classify trust models based on these facts. After this, these schemes are briefly discussed including both positive and negative aspects. Then they are compared using tables to demonstrate their individual characteristics based on a set of factors.

**Centralized Trust Schemes:**
Trust approaches, such as [33, 34, 35] perform computation of trust and reputation in a centralized server. This server gathers information about an event from neighbouring vehicles and then computes the trust of the sender vehicle using a predefined method. These approaches consistently store trust/reputation safely in one place. However, this type of approach suffers from excessive information collection when there is an event and requires sending the revised trust to the respective vehicles which results in undue trust metric dissemination. In some cases, a receiver vehicle asks the centralized server about the trust of a sender vehicle to verify a received message. This augments the driver decision time which is not appropriate for VANETs as prompt decision-making is the most important criteria in VANETs. Otherwise, drivers may cross the junction and miss the detour option to avoid the traffic jam. As a result, congestion would be common on roads if a trust model suffers from a high decision time. In [36], an RSU is solely in charge for the trust verification of the vehicles, and it collects the endorsements/feedback information from vehicles. This is an entity-centric trust model which assesses only entity's trust. Furthermore, the RSU creates, controls, and unites clusters for VANETs. A new vehicle is assigned to an existing cluster upon a joining request when a cluster is already available, otherwise, a new cluster is formed using the only vehicle. This approach divides trust into several components, for example, knowledge, reputation, and experience to make it robust against sybil and wormhole attacks. Additionally, the RSU identifies malicious vehicles, blocks their message, and prevents them from joining another cluster. Though they present a trustworthy cluster, it requires excessive collection of trust metrics, dissemination, and clustering management at the RSUs which requires considerable computational resources. In reference [37], the researchers propose a hybrid trust model which inserts the trust certificate of a vehicle with the event that a receiver utilizes as a weight when computing the trust of data. A vehicle that visits the incident place issues a status update to confirm or deny the event. The vehicle transmits confirmed or denied feedback from the local storage to the RSU when it arrives in a coverage area. Upon reception of the feedback, the RSU sends it to the CA to revise the trust certificate of the concerned vehicle. Afterward, a vehicle contacts the RSU about the revised trust certificate from the CA. This requires further communication with the RSU to transmit feedback and obtain trust certificates frequently which raises a channel availability issue. A theoretical analysis has been carried out to demonstrate the strength of this model considering unreal event broadcasting and an unreasonable trust feedback setting.

**Decentralized Trust Schemes:**
Some trust schemes validate trust in a decentralized manner [38, 39, 40]. Trust is evaluated by the receiver vehicles, and they exchange trust and historical records of interactions with neighbours and RSUs. Most of the blockchain-based trust approaches are decentralized. These approaches offload the computation from the RSU. However, the exchanged messages (direct and indirect trust metrics) can produce sizable traffic in the network in addition to event announcements. A decentralized entity-oriented trust model is typified by [2], where the researchers securely manage assigned credit by a Tamper Proof Module (TPM) on each vehicle.

A sender vehicle first obtains the cost of transmission and the signed message from its TPM. Recipient vehicles consider the reputation of the sender vehicle to trust the message and the trust is adjusted using feedback from all recipients. The analysis shows that this model can thwart false attacks. However, the mechanism for adjusting a revised trust level can lead to too much communication. In [41], the authors compute Packet Delivery Ratio (PDR), familiarity, interaction frequency, and timeliness. They then compute weights for a weight-based aggregated final trust score from the timeliness and interaction frequency. They also examine the time-aware trust of vehicles based on varying recent histories of communication. However, they do not analyse their scheme in presence of any attacker model. In [42], trust is evaluated from experience, trust of the vehicle, neighbouring vehicle information, and the PDR. This approach has a decision manager, trust manager, and route manager. The decision manager decides on the vehicle participation in the packet forwarding function. If this condition is not satisfied, the decision manager notifies a nearby RSU about the mischievousness of the vehicle. The trust manager determines the trust of the path, and the time needed to relay a message to the target. This model chooses a path with the maximum trust and lowest delay. This model is validated in ns-2 and examines the packet delivery ratio, delay, and the number of routes. However, this model only achieves trusted routing. Reference [43] forms a stable trust link graph from the vehicles local trust and then determines the global trust employing a TrustRank algorithm. First, local trusts are computed using Bayesian inference from neighbour recommendations. This local trust data is forwarded to the TA via RSUs to generate the trust link graph. After this, the TA finds the global trust and sends it back to vehicles via RSUs. They also consider driver, vehicle, and behaviour factors in finding a "seed" vehicle and untrustworthy vehicles. The trust value of vehicles is then disseminated from the seed vehicle to the normal vehicles in a Markov process manner using the trust link graph. This approach is compared with one baseline approach to explore its effectiveness in quickly isolating malicious vehicles. Also, it shows that its robustness against newcomer, on-off and collusion attacks is better than the baseline approach. However, this model suffers from the communication overhead associated with the network-wide local trust data collection and global trust data dissemination.

In [44], the data-oriented trust model decides adaptively on received events. Vehicles forward an event message based on the reception of predetermined number of messages about an event or reaching a delay limit between the initial message and the current message received from the RSU. This approach is verified using ns-2 considering false messages, message drop, and message alteration attacks. Furthermore, this is compared against a beacon-based trust model. Nevertheless, the forwarding of event messages is delayed whilst the evaluation takes place. In reference [45], receiver vehicles validate the location and time closeness to an event to establish the trustworthiness and the message newness. This is another data trust model. Then receiver vehicles find the confidence in the received message from each source. Based on a predefined threshold trust score, a receiver vehicle accepts or rejects a message. This approach performs correctly whenever information is received from direct neighbours, otherwise, information from non-neighbours may result in a high false positive rate. In [46], the researchers propose a trust model that evaluates the reliability of the message. This model considers one-hop Emergency Warning Messages (EWM) and multi-hop Event Reporting

Messages (ERM). Recipient vehicles accumulate announcements from in-front vehicles and those vehicles which pass the event location. Vehicles decide about an event based on the time and location closeness of the event and whether it arrives from the in-front vehicle, or if the vehicle has driven through the zone. Then recipient vehicles starts a timer and upon termination, it compares the total positive with the total negative event reports to trust an event. This model is resilient against on-off attacks, bad-mouthing attacks, and sybil attacks. However, this model cannot trace a vehicle due to the lack of identity when disputes arise. The researchers in [47] merge beacons with alert messages to find the status of vehicles. Receiver vehicles match the saved information about a sender with the RSU information whenever they enter a coverage area. This approach splits a digital map into several segments and limits the segments from which a vehicle can receive or reject a message. To send a message, this model verifies a predefined number of endorsements from connected vehicles. However, this model does not perform a security analysis.

A hybrid trust mode is presented in [48] where the receiver vehicle multiplies the judgement of the previous forwarder with its present trust to obtain its judgement about the event and determine whether it forwards the message further or not. An intrusion detection module conducts this activity and estimates the trustworthiness of data. In this model, a trustworthy message can be declined when the trust and opinion of the previous forwarder are low. In [49], a receiver vehicle both verifies the trust condition of the sender vehicle and the legitimacy of the event. Additionally, receivers compare this status with the status of the neighbouring vehicles. After this, a receiver compares its opinion with the neighbours' opinion to decide on the trust of the sender. For instance, if the opinions are equal, then the sender is trustworthy, otherwise, it is mistrusted. Neighbours also exchange trustworthy and malicious vehicle lists to update each other. This approach is evaluated in MATLAB using congested scenarios and the performance evaluation examines the efficacy of the trust computational error and end-to-end delay in the presence of varying vehicle densities. However, the analysis lacks any adversary model.

**Clustering-Based Trust Schemes:**
Clustering is a grouping mechanism that manages trust of multiple vehicles in a group called a cluster. Each group has a leader called the Cluster Head (CH) which coordinates all activities for the cluster. Cluster stability is managed by speed, direction, position, and destination [50]. The CH performs all the communication on behalf of its members. Both intra-cluster and inter-cluster communication are performed by the CH. The CH cannot be malicious so its selection should be trust-oriented as stated in [50]. Another variant of clustering is called platooning where vehicles move together in a group. In [26], a trust-based platoon system is proposed where the users rate the services provided by the Platoon Head (PH). In this way, both good and bad-behaved platoon heads are identified, and the server always advises a well-behaved PH towards the users. The malicious recommendations from users can be isolated using an iterative filtering mechanism to make the evaluation process accurate. As a result, this system always selects the optimal PH for the users. Though a trustworthy cluster can be maintained, but it requires excessive trust metrics collection and dissemination, as well as management of the cluster which demands substantial computational resources. In [51], the authors manage trust for a cluster using the reputations of users. First, vehicles form clusters through a VANET Grouping Algorithm (VGA) under a specific Cluster Head (CH). Vehicles exchange reputation messages inside the cluster. Only the CH updates reputation in the reputation table by checking location closeness, timestamp, and forwarding count. If appropriate, it suggests vehicles for blacklisting to the RSU. Also, a formal proof ensures soundness, completeness, and validation through an inference system. However, this approach is not compared against an existing approach, so it is difficult to assess its performance. In [36], an RSU builds, manages, and merges clusters for the VANET. A new vehicle is assigned to an existing cluster upon a joining request if a cluster is available, otherwise, a new cluster is formed using the requesting vehicle. This approach divides trust into several components, for example, knowledge, reputation, and experience to make it resilient against wormhole and sybil attacks. The RSU also blocks mischievous vehicles from joining another cluster. Additionally, reference [52] designs a trust model for a clustered environment considering the likelihood and impact of making a decision when both the event and the reverse event coexist. The result suggests that the model always chooses a low-risk activity relative to a typical trust-based scheme.

**Reputation-Based Trust Schemes:**
Reputation-based trust models only evaluate the trust of entities. This includes the evaluation of direct interactions and incorporating the indirect trust of neighbouring vehicles in the trust evaluation. Some approaches like [53] perform trust computation at the vehicles and others do the same at the RSUs to manage reputation centrally. In [53], the authors manage the vehicular reputation at the RSUs, and vehicles periodically send all the observed events to nearby RSUs. An RSU then shares revised reputations to each vehicle using the recorded event list. Receivers save all the messages about an event in a decision table until a timer terminates. After this, receivers assess an event as correct or incorrect. Nonetheless, this scheme suffers from high decision time of drivers and communication overhead. Also, the model can encounter data incoherency whilst vehicles are waiting for revised reputation data.

In [17], the entity-oriented trust model employs a false message detection scheme to generate recommendation on the received event which is used with the reputation to compute the final trust. Vehicles first obtained primary and secondary scores from the RSU, and they use these for future communication until the next periodic update. This model is assessed in the presence of fake messages for both city and motorway environments. Nonetheless, this model requires too much trust metric dissemination. In [33], neighbours calculate reputation from past interactions which are stored in a remote centralized server. A vehicle attaches its digitally signed reputation when it sends a message. A receiver accepts the message if it finds the reputation value of the sender is greater than a specific threshold. The authenticity of the sender is verified later, and its reputation is updated on the server. Nonetheless, the centralized server is communicated repeatedly for reputation requests and responses. [34] presents a Reputation-Based Global Trust Management (RGTE) scheme which uses a Reputation Management Center (RMC) to manage reputation centrally. The RMC stores the updated reputation of all participant vehicles in the VANET. Vehicles transmit feedback about their neighbours to the RMC, and it employs statistical central limit theory to disregard unfair recommendations. It revises the reputation scores of vehicles for which it has received recommendations. Whenever a recipient obtains a message, it directly crosscheck the trust of the sender with the RMC. However, this model also suffers from the same problems [33]. Conversely, in [54], the researchers present a reputation system based on opinion piggybacking for VANETs. This approach calculates the confidence of a received event report using majority voting. Furthermore, it uses direct, indirect and a combination of both trust metrics if both are available. However, the approach is susceptible to collusion attack when the longest-serving attacker manipulates the reputation of vehicles.

**Evaluation Side-Based Trust Model:**
As far as we are aware, most of the current trust models employ trust evaluation at the receiver side and this is done only after the reception of events. However, in [1], the researchers present a novel sender-side trust evaluation-based model which employs a Tamper-Proof Device (TPD) to manage trust manipulation of multiple drivers. This model employs a withhold period to allow other drivers to report the original message if any receiver believes the sender sent an untrue message. In this case, the sender does not obtain the reward rather it waits for the decision from the RSU to get the revised trust of the current driver using the reward/punishment issued by the RSU. When an untrue attack message is disseminated, an RSU takes the responsibility to resolve the dispute using a weighted voting-based method from the clarifier's trusted feedback to determine the benevolent and malicious activity. Here, a reporter vehicle is the vehicle which generates an untrue attack report in the VANET. A clarifier vehicle is a regular vehicle that transmits feedback or response to the queried RSU about a conflicting event when they visit the event location. When the conflict is resolved, the RSU then sends reward and punishment messages to the corresponding drivers. An extension of this RSU judgement scheme is also provided in [55] where fuzzy RSU judgement is assigned based on Driver Past Behaviour (DPB), severity of the event, and RSU confidence about the sender or reporter driver. Otherwise, if the withhold period expires with no challenged event received at the sender vehicle, then the TPD adds the reward to the current trust of the driver. This model thwarts untrue attacks, inconsistent attacks, and cooperation attacks and the validation is conducted for several urban scenarios with different traffic conditions in Veins.

**Blockchain-Based Trust Schemes:**
Blockchain-based trust models adequately improve security and trust management in VANETs. Many trust approaches use blockchain as a key repository of tamper-proof information for trust management [35, 38, 39, 40, 56, 57, 58, 59, 60, 61, 62, 63]. The key benefit of using blockchain is the ability to maintain consistency of information by sharing the same copy of the blockchain among the RSUs. Every block in the blockchain is managed and updated by the RSUs. However, during block insertion, a fake block can be inserted by a fake RSU. Consensus mechanisms, assuring block verification and validation, demand more research as suggested by [64]. More vehicles are increasingly seen on the road which influences to use of multiple blockchains to fit this demand. [64] suggests designing a suitable framework to address the scalability issue. It is also common as suggested by [64] that some models run miner selection at RSUs, whereas, in other models, the vehicles run the miner selection algorithm. As yet, there is no standard which defines the entity which should perform miner selection in the blockchain.

Reference [38] presents a decentralized reputation approach based on Secure Multiparty Computation (SMC) and blockchain for VANETs. In SMC, each user inputs a private number to the shared agreed function and then expects an output from the function. In this approach, individual assessments from vehicles are kept private and the resultant reputation data is made public. The blockchain provides network-wide consistent reputation data for vehicles. An analysis is conducted considering collusion attacks. However, it is not compared with any baseline approach. [58] employs a multi-level blockchain for trust management in VANETs. This model consists of three parts; the first one employs a horizontal trust mechanism at each vehicle using support vector machine, k nearest neighbour, or random forest. Then vehicles notify classification decisions (normal or malicious neighbours) to RSUs. The vertical trust scheme applies a verification algorithm at the RSUs to distribute the trust list. Third, a Distributed Trust Management (DTM) scheme allows the RSUs to utilize the blockchain to share the trust list and to determine the vehicle class. This model is implemented in Veins and can thwart sybil attacks using the VeReMi dataset. However, trust formation, aggregation, composition, propagation, and updating require cooperation at various levels. Reference [59] proposes another blockchain-enabled distributed trust-based location privacy protection scheme for VANETs. Dirichlet distribution is employed for trust computation between the requesting and cooperative vehicles. After an evaluation, the result is transmitted to a nearby RSU to generate a block to insert into the blockchain. The RSU transmits this result to neighbour vehicles and shares it with other RSUs in the VANET, so the information remains trustworthy. A distributed k-anonymity scheme is employed to construct a trust-based anonymous cloaking region. The trust in the vehicles is evaluated by using the historical trust records and present behaviours. This model is implemented in JAVA and the blockchain is deployed in HyperLeader. The security analysis suggests the approach is resilient to bad-mouthing attacks, on-off attacks, whitewashing, and sybil attacks. Furthermore, this is compared with a baseline model and demonstrates the privacy leakage rate reduces gradually over time. However, the malicious vehicle detection rates demand at least twenty to thirty rounds to achieve high accuracy although they consider this as the initialization phase. In reference [35], a blockchain-based trust model is also presented to preserve the privacy of VANETs. Vehicles utilize the certificate as a pseudonym to access a Location-Based Services (LBS) while preserving the privacy of a real user. A Dirichlet distribution-based trust management algorithm is also developed to assess vehicle behaviour while the Certificate Authority (CA) manages trust records from users securely into a blockchain centrally. An RSU-dominant algorithm is also developed to construct the cloaking region for privacy protection. A security analysis and trust progression in the presence of on-off attacks and bad-mouthing attacks is examined. This is a better privacy protection scheme than [59], but it suffers from latency.

Reference [60] presents a data-oriented Blockchain-Based Traffic Event and Trust Verification (BTEV) framework. The framework handles security, trust, and privacy through a two-stage verification of events and a two-phase transaction for rapid event notification. This framework can prevent selfish attacks and false event relaying. Nevertheless, the two-phase Proof of Event (PoE) consensus yields latency when traffic densities are high and when vehicles request bulk messages, which may congest the network. Also, a blockchain-based hybrid trust model is presented in [39]. In this model, receiver vehicles compute the trust of received messages and then believe or decline them based on a limit. Additionally, a receiver evaluates the trust of the sender vehicle and transmits the revised trust at times to an RSU which combines these values to estimate the reputation of the sender vehicle. The RSU first determines the reputation of the message sender and packs these reputations into a block and then operates as a miner to insert it to the blockchain. This model is examined in OMNeT++ and compared against a decentralized trust model to demonstrate it performs better in the identification of malicious vehicles and dropping of malicious messages. Nonetheless, with a high number of malicious vehicles, they can assign higher trust levels to themselves so that the malicious message identification rate may fall.

In [62], the researchers present a hybrid, three-layer blockchain-based trust model which is based on reputation regression, Dirichlet distribution, and punishment revocation. Blocks are saved in the cloud and a CA manages the key sharing of vehicles and the registration of RSUs and vehicles. This model considers simple, strategic, and slander attacks along with both normal and malicious servers and reviewers in their analysis to demonstrate the precision and recall rate are better than a baseline model. Nevertheless, this model does not employ any reward mechanism to encourage benevolent behaviour from vehicles. In reference [40], the researchers propose a distributed blockchain-based trust model which selects a message assessor through RSU cooperation. The model finds a score for messages, the sender, and the assessor vehicles. Then, they determine the global trust score of a vehicle based on the score and message quality. They store trust metric in a blockchain and use a consensus mechanism to add blocks. They claim that this model can thwart sybil attacks, ballot-stuffing attacks, message spoofing attacks, and bad-mouthing attacks. The average delay in their consensus mechanism is compared with two other baseline models. The researchers in [56] present a blockchain-based intelligent trust model that validates the received event using Bayesian inference. Every vehicle issues a score for each sender vehicle which the RSU uses to produce a block of trusted vehicles. The block is added to the trusted blockchain by an RSU. Nonetheless, a compromised RSU can build a fake block using untrusted vehicles and insert it to the trusted blockchain.

**Functional, Recommendation and Role-based Trust Schemes:**
Some Models apply functional trust, recommendation trust, role based or experience-based trust to obtain the final trust. Functional trust is defined by how likely a vehicle is doing its activities. Recommendation trust is the quality of recommendation from a vehicle about other vehicles. Role-based trust indicates that how a vehicle is performing its assigned role in a network. The researchers in [65] first check the trust of data and then compute the trust of a vehicle from the functional and recommendation trust. The investigation considers simple attacks, bad mouth attacks and zigzag attacks, and determines the precision and recall with truly malicious vehicles to examine the accuracy. However, this model also suffers from latency. Alternatively, a self-organizing hybrid trust scheme is presented for both city and countryside scenarios [66]. This model maintains previous experience and then validates the received event by assigning a credit score. This scheme computes the trust for each distinctive message to accept the message with the maximum trust for a specific incident. This model can identify false source locations, incident locations, and incident time, and can withdraw messages from mischievous vehicles. Nonetheless, this model is not compared against a baseline. In [67], the researchers combine experience-based, role-based, and majority-based trust to compute the aggregated trust score. Whenever a receiver obtains a message, it looks up a local matrix sorted by role and experience to find the trustworthiness of the message. Neighbours are saved and ordered using their contributions to formulate their opinions. Nevertheless, this model considers frequent meetings between vehicles which is inappropriate for VANETs. In reference [68], a vehicle applies cognition to build contexts around an event to conclude trust. It constructs a context using an ontology which relates a group of interlinked concepts (for instance, event, vehicle, and evaluation). This model considers opinion, experience, and role for the trust verification. In outlier detection, this model uses speed, time, and distance thresholds. Furthermore, this model finds the confidence of the report besides finding the trust level for every report. The model is analysed in MATLAB using both city and countryside scenarios and compared against baseline trust models using a confusion matrix. However, mischievous vehicles can still avoid the outlier-based malicious identification and transmit false events within the permitted threshold.

**Active Detection-Based Trust Schemes:**
Active detection is a method where a vehicle sends a message to a neighbouring vehicle to forward it to an RSU or other entity. If the neighbour relays the message as the sender sends, the relayer is considered trusted. In active detection, a vehicle observes a relayer for a specific action. An entity-oriented trust model is proposed in [57]. The researchers employ an active detection method and use blockchain to manage trust for VANETs. Active detection strategy is used as follows: vehicles send probe messages to neighbours to evaluate their direct trust by requesting them to forward these messages to nearby RSUs. They then wait for the response from the RSU. Vehicles also gather reference trust of other vehicles with usual behaviours from the locality to relay data using only highly reliable vehicles. After this, vehicles send an updated trust list to the RSU to store it to a blockchain. The RSU disseminates trust data to vehicles and pedestrians. This approach is evaluated to determine the accuracy and error rate as well as being compared against both centralized and decentralized approaches. However, the active detection process demands extensive probe packet generation. In [63], the researchers propose a machine learning and active detection-based trust model to assess the trustworthiness of vehicles and events. Active detection assists in discovering the indirect trust of neighbours and a Bayesian classifier is employed to detect a malicious vehicle. Here, a receiver vehicle discovers the trust of the source vehicle by multiplying the direct and indirect trust and matching the result against a threshold to accept or reject a message. Blockchain is employed to preserve the trust and certificate of vehicles. It accomplishes Vehicle to Vehicle (V2V) authentication using a smart contract on the blockchain. This model is evaluated in Python and observes the trust score progression with different indirect trust levels and different message accuracies. The model identifies mischievous behaviour at a fixed time from specific vehicles. Nonetheless, every active detection demands two additional messages for each evaluation.

Furthermore, [69] proposes a hybrid trust evaluation scheme using both direct and indirect trust and accepts messages from vehicles when the trust score is greater than a threshold. A Gaussian Naïve Bayesian classifier is trained to predict the probability value to use it as the direct trust of vehicles. Active detection is used to compute the indirect trust of vehicles by requesting a neighbour vehicle to forward a message to the RSU and also the sender directly sends this message to the RSU. The RSU then compares these two messages to determine whether the neighbour forwards an altered message to determine whether it is malicious or not. The RSU then updates the neighbour vehicle's indirect trust into blockchain. It stores both indirect trust and certificates into blockchain. Additionally, it employs smart contracts to extract identity information and trust values from the blockchain. It uses V2V authentication and digital signatures to maintain the integrity of the messages. This approach also evaluates an event using trust of vehicles and distance from the indicated event location. After this, it computes the probability of the event to confirm its reality from the responses of the nearby vehicles. Blockchain confirms the transparency of the data throughout the network. However, the scheme suffers from communication overhead as it uses active detection based malicious vehicle detection and collects recommendations about events from vehicles.

**Artificial Intelligence-Based Trust Schemes:**
Some approaches, for example, [63, 70, 71, 72] employ Artificial Intelligence (AI) to perform the trust computation, for instance, deep learning [73], reinforcement learning [28, 74, 75], decision trees [71, 72] are used for trust management. Furthermore, fuzzy logic [28, 70] is used to handle uncertainty in VANETs. However, the application of these techniques has some limitations. For example, in [73], deep learning is used by the RSU to verify a sender and its messages, and the sender's trust is computed by the trusted entity. When receivers receive this event, they need to wait for verification from the RSU. Consequently, the driver decision time increases which can cause them to be queued around an event or they may bypass a possible detour junction as they are moving. This increases driver frustration. Furthermore, decision tree-based approaches [71, 72] to date only applies in detecting fake position attacks from basic safety messages. The application of reinforcement learning is limited to finding a factor or adjusting a factor in relation to the trust computation. Therefore, the application of machine learning in trust management for VANETs has so far failed to achieve the VANET strict timing requirements and/or suffers from communication overhead. However, we believe careful implementation of these algorithms can improve performance.

The researchers in [71] present a data-oriented machine learning (decision tree, KNN, random forest, and Naïve Bayes) based trust model to detect location spoofing attacks from sequential Basic Safety Messages (BSMs). Vehicles transmit BSMs to an RSU which employs a detection framework using the stored data and the received BSMs. The model is trained with a VeReMi dataset which includes both legitimate and malicious data that helps to classify future information. Their analysis investigates the accuracy, precision, and recall of each machine learning approach. Results suggest that the KNN and random forest demonstrate better accuracy in identifying attacks. However, this only examines BSMs data to identifying false position attacks. Reference [72] presents an Explainable AI (XAI) system for trust management of autonomous vehicles through an ensemble learning algorithm and a decision tree-based model to separate malicious vehicles from benevolent ones. Explainable AI makes the complicated machine learning and deep learning model more comprehendible. The simulation of this model suggests the suitability of this model for VANETs in terms of accuracy, precision, and recall. However, this approach only analyses fake positional data to identify malicious vehicles using the VeReMi dataset. Reference [75] presents a trust model to evaluate the trust of vehicles with high accuracy in the presence of a high percentage of mischievous vehicles. The model verifies an event via a coefficient-based weighted method from the external and sensed information as well as self-experience (internal information). The final trust is then checked against a threshold-based trust to decide whether the event is bogus or legitimate. Furthermore, a reinforcement learning algorithm is employed to adjust the trust assessment function based on prior results. Vehicles which comply with the protocol are treated as normal and those that go against the protocol either purposely (malicious) or inadvertently (defective) are treated as adversaries. The trust model is investigated in Veins using real-life scenario data and the precision is compared against Bayesian, DST-based, and voting-based models by changing the influence of the fake information. In an unexplored road situation, when there is no internal information available, if there is more malicious information than reliable information this model may reach a biased decision about an event. As this model requires the accumulation of external information to compute the trust of an event, it may delay the evaluation process until a fixed period has elapsed or a certain number of reports have been received.

Moreover, fuzzy logic [28, 70] based trust mechanisms require repeated sensing of messages from neighbours. In [76], the authors propose an infrastructure-based trust scheme which considers the severity level of the safety events. This model uses three fuzzy sets to decide on a received message. The decisions are based on trust thresholds which are rejecting and dropping the message, accepting but not relaying to others, or accepting with relaying to others. Moreover, this model evaluates the reputation of the sender vehicle using direct trust, and indirect trust from neighbours and RSUs. This evaluation applies weights to each metric. This model is simple, accurate, fast, scalable, and resilient to some threats. However, the scheme lacks an estimation for communication overhead, identity, and privacy management. [28] also utilizes fuzzy logic to compute direct trust and Q-learning to determine indirect trust to determine the final trust. This model examines precision and recall metrics in the presence of varying numbers of mischievous vehicles. However, the communication overhead is high as it needs regular sensing of hello messages from the neighbours. The authors in [70] also employ fuzzy logic to calculate the relaying and coordinating trust. Then the final trust is computed from these two and a trusted route is explored using a set of rules and experiences. However, this scheme only achieves trusted routing to forward a message along the most trusted route.

In reference [77], receivers use a logistic regression-based hybrid trust model to detect misbehaviours in the VANET. The logistic regression iteratively progresses based on the trust score to correct for misjudgements. The trust of the sender vehicle is revised using the received messages initiated from them. In some cases, when an observation is not present, the receivers utilize learned events from trusted sources to confirm the received message and assess the trust of the sender. The receiver disseminates a list of malicious and honest vehicles to neighbours. This model is validated in OMNeT++ using a circular route to redirect vehicles iteratively. First, they examine the trust progression in the presence of on-off attacks. The model is also compared with weighted voting and a majority voting model and analyses accuracy to demonstrate the efficacy of detecting malicious vehicles. However, a malicious vehicle can benefit by changing an attribute, for example, acceleration, braking, or location data in the Basic Safety Messages (BSMs). In [73], the authors propose a deep learning-based hybrid trust evaluation model to thwart internal attacks. The deep neural network contains four hidden layers and computes rewards based on driver behaviour and classifies trustworthy or malicious behaviour using another deep neural network. The trust of a vehicle is computed from reward points using a three-hidden-layer deep learning scheme by the receiver vehicles where messages are classed as deceitful or trustworthy by an RSU using a two-hidden-layer deep neural network. This model is validated in ns-3 using the open-source library TensorFlow 1.6.0 and then compared against two other baselines considering the computational overhead. However, the model suffers from an inefficient trust update mechanism as it demands a chain of communications to the top-level authority of the architecture from the vehicle for each message dissemination.

**Software-Defined Networking-Based Trust Schemes:**
Software-Defined Networking (SDN) enables configuring and management of networks using software rather hardware. In VANETs, machine learning algorithm and SDN combinedly used to achieve trust management. In [74], the researchers present a software-defined Trust-based Deep Reinforcement Learning Framework (TDRL-RP) for VANETs. This entity-oriented trust model computes the trust based on the packet forwarding performance of neighbours and a SDN-based agent selects the highest trusted path using a Convolutional Neural Network (CNN). Additionally, in this model, the VANET routing selection is modelled as a Deep Reinforcement Learning (DRL) problem where the objective is to determine the most trusted routing path from the source to the destination. A deep Q-network followed by a CNN model is employed which takes the state as the input for training the network and generates a Q value as the output. This approach is analysed in OPNET by using TensorFlow and compared against the AODV routing protocol and an existing SDN-based approach. The results confirm that the approach exhibits a better packet forwarding rate and network throughput. Nonetheless, the trust model of this approach is used only in selecting a trusted route to a destination.

**Probability and Statistics-Based Trust Schemes:**
In trust management, both probability and statistical techniques are widely used to compute direct trust, indirect trust or the final trust. Two of the most popular probability and statistical methods are Bayesian inference and Dempster-Shafer Theorem (DST) which are used by many researchers for trust management in VANETs. For example, the entity-oriented model in [78] uses the DST to combine direct and indirect trust to find the final trust of vehicles. Also, in [79], the researchers employ DST to merge data from several neighbours about an emergency event. In this model, vehicles confirm the location of an event from beacon messages. In their evaluation, they also consider message alteration, suppression, and bogus attacks to evaluate the trustworthiness of the model. However, this model also results in delay as it accumulates messages from neighbours. An alternative DST-based model described in [80] uses old beliefs from neighbours which may generate wrong results because of using inaccurate trust metrics from neighbours. In [29], DST is used to combine many independent beliefs to find the recommendation trust. Also, they use Bayesian rules to compute the direct trust of vehicles in VANETs. The researchers in [81] use a Bayesian filter and watchdog mechanism to find the direct trust and then the final trust from the weighted combination of direct and indirect trust. Also, [82] uses a Bayesian criterion to determine the probability of a message being malicious or not from a specific sender. Using this technique, it finds the suspicion level and then the trust level of the vehicle or driver. The researchers in [81] use a Bayesian filter and watchdog scheme to determine the trust of vehicles. Vehicles communicate with an RSU to obtain the indirect trust of the target vehicle. After that, a vehicle sets the final trust from a weighted combination of direct and indirect trust. Based on this score, vehicles are marked with a state from: malicious, heavily suspicious, lightly suspicious, and normal. Although they compare this model against a baseline considering accuracy and errors, they do not include a punishment mechanism. Also, [29] is an entity-oriented trust model that uses direct trust and indirect trust where a Bayesian rule is used to compute direct trust, and the Dempster-Shafer Theorem (DST) is used to obtain the recommendation trust. This model aggregates many separate beliefs about a vehicle to compute its overall trust. Nonetheless, an incorrect recommendation can incorrectly influence the trust evaluation.

The researchers in [82] present a probabilistic trust scheme to evaluate the trust of received information. This model finds the distance and geolocation from the Received Signal Strength (RSS) of the messages. It does not convey a trustworthy message if it is beyond a specific distance. A Bayesian Inference-based voting mechanism is presented in [83] which assigns a time parameter to each road segment and event messages carry the road ID when they are conveyed [83]. Every vehicle keeps a local map and a database to revise these records. Receiver vehicles recompute a new route using the Dijkstra algorithm upon receipt of a message. However, this demands regular maintenance of a local database. In [84], the researchers present a dynamic hybrid trust model which assigns a weight based on the role of the vehicle and the type of application. The dynamic entity-oriented trust model prevents selective forwarding attacks and black-hole attacks by marginally sacrificing the performance of a Greedy Perimeter Stateless Routing (GPSR) protocol. The data-oriented trust model finds relations among data and computes trust based on traffic patterns and utility theory. This study examines the impact of a trust model on a routing protocol and compares this to a GPSR routing protocol. The data model can be further improved by choosing proper utility parameters. In [85], the authors propose a trust model which uses Bayesian inference to determine the direct trust and recommendations to determine the indirect trust for VANETs. The direct trust calculation is based on penalties and time-decaying information. Moreover, the confidence of direct trust is verified against a threshold score to avoid a costly recommendation trust calculation. The approach achieves more successful interactions than two baseline approaches. However, the investigation only considers packet drop and interception as malicious behaviours.

**Data Collection Nature-Based Trust Schemes:**
Trust model data collection methods differ in the number of hops from the source to the data collector. They can be classified into three main categories which are direct [57, 45, 86], indirect, and hybrid [28, 78, 81, 76, 87, 54] using the nature of the data collection [78]. Methods that only rely on direct trust data, collect information from the one-hop neighbours. Indirect trust data is collected from the recommendation of one-hop neighbours about the non-neighbouring nodes. A solely indirect trust-based model is rare in the state-of-the-art, but this concept often forms part of a hybrid data collection method. These models evaluate trust by collecting trust data which increases communication overhead and driver decision time in some instances. In [86], the researchers use an ant colony optimization algorithm as well as incorporating direct experience with feedback information to compute the trust of data. However, this data-oriented trust model introduces longer delays since it requires time to collect, analysis and distribute data. Some factors that impact this increased latency are the distance between the nearby RSUs, traffic densities, traffic situation, and a vehicle evidence threshold. Consequently, data delivery latency varies from zero to thirty seven seconds based on their experimental scenarios. Also, the frequent communication between RSUs and vehicles affects latency.

In [88], a hybrid (collecting both direct and indirect data) Analytical Hierarchy Process (AHP) based trust computation scheme is proposed. In this entity-oriented trust model, a Perron-Frobenius theorem-based direct trust, a certification-based indirect recommendation, and reputation are used to compute the trust of a vehicle. An implementation scenario is presented to analyse the communication delay between vehicles though this is not a real-world scenario. Reference [89] achieves reliable data delivery and presents an intrusion detection module to thwart Denial of Service (DoS) attacks. This scheme calculates the trust of neighbours using a sender's direct trust data, indirect trust/opinion received from the previous relayer, weight of the official vehicle, and prior verified sender data. In this approach, a routing protocol is used to forward packets along the most trusted path. However, this approach only suggests a trusted route for packet forwarding besides thwarting DoS attacks. A data trust model for VANETs is proposed in [78] where an RSU computes the direct trust of the event by determining the similarity between the beacon and the received event and then notifying nearby vehicles. When indirect trust is available, both direct and indirect trust are aggregated using DST. This approach is validated in the presence of message alteration and false message attacks to evaluate the accuracy, f-measure, precision, and recall. This model is also compared against two baseline approaches. However, this model suffers from latency whilst the RSU decides and shares its judgment about an event.

[87] admits or denies a new vehicle from the direct and indirect trust calculation using a historical security vector of events. The security vector is first calculated based on past behaviours and then the Authority Unit (AU) is contacted to obtain an historical security evaluation to determine the direct trust of the vehicle. For the indirect trust computation, the vehicle recommendation trust vector is constructed from neighbouring vehicles. This hybrid trust model also utilizes the correlation coefficient to discard malicious recommendations. However, this model does not use any adversary model while assessing efficacy. Reference [18] determines the trustworthiness of messages considering only direct trust. Every vehicle saves past interactions and the trust of all neighbouring vehicles. This model identifies eavesdropped messages and false events. Nevertheless, they do not consider false trust messages from malicious vehicles. To improve the correct decision of vehicles, a hybrid trust model is presented in [90]. This approach both considers beacons and emergency event dissemination.

The status of an event can change from 1 (event exists) to 0 (event does not exist). The role of the RSU is to deliver back and forth trust certificates between a vehicle and the TA. The TA updates trust certificates of vehicles. First, this model finds the likelihood of sending a message from one vehicle to another using a likelihood function. Then the direct trust from one vehicle to another is determined using a Beta distribution function. The final trust of a vehicle is obtained from the weighted sum of direct trust and an historical record placed within the trust certificate. Also, this approach considers different states for example, idle, collection, decision-making, and update. For example, as a vehicle progresses towards a traffic event it changes its state from idle to the update state. This model considers false messages, blackholes and feedback attacks as adversaries. Reference [91] presents a trust model which can operate across various traffic densities. This model computes the trust of neighbours, detects malicious vehicles with/without RSUs, and finds the most trusted route to transmit a message. This model is compared with a routing protocol using black hole attacks as an adversary model. It is also verified without RSUs, using direct and indirect trust, and there is a scope for improved robustness considering other adversaries.

**Privacy-Preserving Trust Schemes:**
In the state-of-the-art, it is found that some trust models deal with other security requirements besides trust management. These trust model enforces privacy of drivers and vehicles using pseudonyms and cloaking region. For example, in [59], a distributed k-anonymity method is used to construct a trust-based anonymous cloaking region to protect privacy. Cloaking region-based privacy protection achieves location privacy by hiding users' exact location within the cloaking region. Also, in [35], vehicles use pseudonyms to access the Location Based Service (LBS) while protecting the privacy of vehicles. To protect users' identity information, a privacy protection-based communication system is proposed [92]. This scheme uses identity-based cryptography and Elliptic Curve Cryptography (ECC) to achieve the stated objective. In [93], a weighted voting-based data trust model is proposed for VANETs. This model assigns a lower trust score to distant vehicles than to nearer ones considering an event location while vehicles evaluate an event. Additionally, it employs a privacy scheme which employs a pseudonym-based public key. However, this model considers distance as the only metric for trust manipulation.

An adaptive trust-privacy framework consisting of an Adaptive Link-ability Recognition Scheme (ALRS) and an Adaptive Trust Management Scheme (ATMS) is presented in [94]. ALRS preserves privacy by hiding identity and supports trust management by exposing the identity and vice versa. ATMS checks data from other vehicles and revises reputation. Nonetheless, the hybrid trust model suffers from prejudicial decisions when a malicious vehicle rating is very high. Reference [95] determines the trust of other vehicles using a vehicle's daily highest operating time. This model monitors the operating time of normal vehicles and maintains a track of valid and invalid messages transmitted from vehicles and revises the trust details consequently. Trust assessors keep their database updated by disseminating the latest information and vehicles receive revised trust information from them. A message from a vehicle contains a signature, trust details, and a timestamp. Receiver vehicles first check the received message by assessing a hash of the trust and timestamp and then find a confidence score to accept or deny it. This hybrid trust model is evaluated in ns-3 and a theoretical analysis is additionally presented to ensure that the model meets authentication, integrity, privacy, and non-repudiation requirements. Nonetheless, a malicious vehicle can drive for a longer period to become a trusted assessor, and the evaluation does not consider any known adversary in their validation.

**Cryptography-Based Trust Schemes:**
Trust models consider many symmetric and asymmetric key cryptography algorithms for security in VANETs. Besides these, digital signature, hashing, identity-based cryptography and batch signatures are also used. In reference [96], the researchers use Hash Message Authentication Code (HMAC) and digital signatures to manage the trust of vehicles. RSUs evaluate the vehicle trust based on neighbour trust values and rewards. The scheme also measures the communication overhead for a number of vehicles. The verification can use ID-based and batch signatures, and the trust computation can also determine punishments. The researchers in [19] also propose an ID authentication and symmetric HMAC-based trust management approach. The receiver vehicles verify the trust of the sender vehicle every time a message arrives. However, the scheme is not validated on any simulator. [97] combines a reward-based trust scheme with hybrid cryptography to secure the VANET. The TA sets the trust of newly registered vehicles, and it updates the trust of the sender vehicles centrally for every message broadcast. The receiving vehicles forward messages to the RSU to verify their authenticity and the integrity of the trust of the sender and then checks the trust threshold to accept or reject them. This approach assigns different reward points as per the severity of the event. The severity of an event is determined by its effect on human lives. However, the approach does not provide detection of false alarms. The researchers have conducted a theoretical analysis and evaluated its efficacy against a baseline. The trust verification of each message initiates communication among the entities from various levels in the hierarchy of the architecture (first RSU to the Agent of the Trust Authority (ATA), and then ATA to the Regional Transport Office (RTO)). This results in considerable communication overhead and adds latency to the decision-making process.

**Trust Schemes-Based on 5G Enabled VANET:**
In VANETs, vehicles, RSUs and the TA are common elements where vehicles and RSUs communicate using the DSRC protocol. As DSRC suffers from scalability, limited capability, and sporadic connectivity, 5G networks can provide better service to achieve reliable communication and low latency [61]. A scheme using a 5G base station and 5G Subscriber Identification Number (SIM) for each vehicle is proposed in [61] to address the trust of real-time traffic incidents. 5G base stations provide wireless Internet access to the vehicles. In this model, vehicles upload traffic situation videos with a tag to a cloud server which nearby vehicles score based on whether it is accurate or not. An RSU serves as the DSRC access point for the vehicles. Also, Proof of Work and Proof of Stake-based blockchain is used. The message is validated first to compute the trust of an event, and the legitimate entity uploads the video into the cloud with blockchain. Malicious traffic event broadcasts are dropped, and malicious vehicles are barred. This approach is evaluated in OMNeT++ using the crypto++ (https://cryptopp.com) library and a security analysis is conducted in the presence of malicious vehicles and an RSU. The trust can be determined accurately in the presence of a low density of malicious vehicles but when it is over 5%, disparities arise. A three-layer 5G-enabled trust evaluation architecture is proposed for intelligent transport system [98]. The layers are the application layer, an edge trust evaluation layer and the intelligent terminal layer. The RSU collects data from the intelligent terminal which are evaluated by the edge servers. Federated deep learning is used to compute the trust of users. Trust evaluation considers hierarchical rewards and punishments. Additionally, a blockchain is used to verify trust and for storage purposes.

**Fog Computing-Based Trust Schemes:**
Fog computing or fogging or edge computing brings cloud computing services closer to the end-users to achieve efficiency in data handling, storage, computing and communication. The edge devices are called fog nodes which are used to enhance security applications. It reduces the data volume by not sending to the clouds and hence it requires processing of some data locally. Thus, it enforces security and can achieve data privacy. In [99], the researchers use fog nodes to collect and filter the trust records. This approach classifies vehicles as either frequent or occasional visitors. This approach isolates the reputation of safety-related tasks from the non-safety tasks. It is simulated in MATLAB and compares the message overhead against an experience-based trust model. The results suggest that the model reduces the message overhead and offloads the computations to the infrastructure from the vehicles. However, the fog nodes and RSU need considerable communication to obtain the updated trust. Additionally, the researchers in [100] determine the accuracy of location using fog nodes. This model employs fuzzy logic to compute the trust from experience, plausibility, and location accuracy. It can thwart bogus attacks and message modification attacks. Nonetheless, vehicles contacting with edge devices for location accuracy raise the communication overhead.

**Social Network and Email-Based Trust Schemes:**
A Vehicular Social Network (VSN) maintains social relationship on top of a vehicular network [101]. There are two types of VSN exist which are centralized and decentralized. In centralized system, vehicle and driver's relationship is stored in a central cloud. In contrast, in decentralized VSN, vehicles share and maintain their relationship with neighbours. In email-based trust, the frequency of email communication between two users is used as a metric for trust computation [102]. Reference [102] presents a social network and email-based trust framework for VSNs. VSN consists of VANET and Online Social Network (OSN). This functional architectural framework provides value added services and applications for drivers and passengers. Through this hybrid framework, the researchers reduce the gap between the entity trust model and the data trust model. However, the framework is not analysed with a real-life traffic scenario.

**Infrastructure-less Trust Schemes:**
Some trust models compute trust using RSU like infrastructure and others do not employ any RSUs. The trust models which do not use any infrastructures for trust management are infrastructure-less trust model. Reference [103] presents a data-oriented trust scheme which computes trust from content similarity, content conflict, and route similarity. It assigns an individual trust score to each event which reflects the probability of being true. It resolves conflicting situations by giving importance to higher trust scores. Nevertheless, activity monitoring is restricted in this approach as there is no RSU. Reference [104] presents the FACT framework consisting of two modules for achieving reliable data delivery. The first module verifies the security of the event and assigns a trust value to every road segment and neighbourhood. If an event is reliable, then the second module finds a highly trusted route for forwarding the message. Nonetheless, this application-centric approach does not support authority monitoring.

**Evaluation Nature-Based Trust Schemes:**
Most researchers classifies existing trust models based on how they evaluate trust in VANETs. Some models only evaluate an entity whereas others evaluate only data. Additionally, many trust model evaluate both entity and data for VANETs. Based on this fact, they can be classified into three classes. The first one is the Entity-Oriented Trust Models (EOTM) which only evaluate an entity's trust. Trust models proposed in [33, 38, 39, 36] follow this evaluation approach. The second class is the Data-Oriented Trust Models (DOTM) which verify the reliability of data. Some well-known many data trust models are proposed in [71, 75, 78, 79]. Finally, Hybrid Trust Models (HTM) verify both an entity's trust and the credibility of the data. Few examples of hybrid trust models are [77, 84, 85, 94]. This is a well-established classification system followed by most researchers when considering trust models for VANETs.

## COMPARISON OF THE STATE-OF-THE-ART TRUST MODELS

In this section, some selected trust models are compared using tables to illustrate their differences considering basic principles, trust metrics/feedback collection, and trust evaluation. Tables 2 and 3 summarize entity trust models, Tables 4 and 5 summarize data-oriented trust models. Tables 6 and 7 summarize hybrid trust models. In addition, this comparison considers the different types of adversaries each trust model can thwart as well as what analysis was conducted to validate each trust model. Furthermore, whether an approach is compared with a baseline or not is identified, and what traffic scenarios a trust model considers during the evaluation is also noted. However, most of scheme adopt a receiver-side approach. Hence, they suffer either from communication overhead or from decision latency and some approaches suffer from both. The reason for this is explained in Section 6. In the following tables, E indicates the scheme is an Entity type, D is a Data type and H is a Hybrid type.

**Table 2: Comparison of Entity Trust Models (Part 1)**

| Ref. | Type | Any Road-side Units? | Underlying Principle | Feedback Collection | Trust computation | Adversary Model | Analysis | Baseline Comparison | Traffic Scenarios | Simulators |
|------|------|---------------------|---------------------|---------------------|-------------------|-----------------|----------|--------------------|--------------------|------------|
| [2] | E | No RSU | ✓ TPM stores credit scores to send or receive messages.<br>✓ To send a message, the transmission cost is determined, and it is paid back for sending a true event. | Each receiver sends an acceptance or refusal notification to the sender. | ✓ Majority opinion from the feedback used to find the reward.<br>✓ A reward is added to the credit account of the sender vehicle. | ✓ Alteration attack.<br>✓ False message<br>✓ Selfish behaviour. | ✓ Malicious vehicle detection delay and false positive rate.<br>✓ Reception rate of corrupted data and reception ratio in the presence of selfish nodes. | Node detection percentages (in the presence of varying rates of malicious vehicles) are compared theoretically. | Urban and highway. | ns-2, VanetMobi-Sim, SUMO. |
| [17] | E | RSU | ✓ Evaluate the accuracy of information from feedback and | It generates feedback. | The source gets a score based on the information accuracy. | ✓ False message. | Show the deviation between the actual and estimated accuracy for different scenarios. | No baseline comparison. | Urban and highway. | Veins, OMNeT++, SUMO. |

| Ref | | | | | | | | | |
|---|---|---|---|---|---|---|---|---|---|
| | | | remove malicious feedback.<br>✓ It blacklists vehicles. | | | | | | |
| [28] | E | No RSU | ✓ Fuzzy logic considers cooperativeness, honesty, and responsibility to find direct trust.<br>✓ Q learning evaluates the indirect trust. | Repeated sensing of "hello". | Trust is calculated from the direct and indirect trust. | ✓ Bad mouth attack. | Analysis of:<br>✓ Precision and recall.<br>✓ Packet delivery ratio under different malicious rates. | Compared against a deterministic trust and a without trust-based scheme. | The freeway has two lanes in each direction. | ns-2 |
| [33] | E | Access points | ✓ The reputation score determines the reliability of an event.<br>✓ The reputation server revokes the reputation of a malicious vehicle when a specific condition holds and does not issue any certificate. | Feedback | When the time-decaying reputation score of the sender vehicle is greater than a threshold, the receiver vehicle accepts the message. | Theoretical proof against:<br>✓ False message.<br>✓ Reputation manipulation attacks. | Analysis of:<br>✓ Message drop rate versus access point distribution.<br>✓ Unavailability of access point and reputation server. | Compared with the two baseline approaches. | Urban scenario taken from Pittsburgh. | GrooveNet |
| [38] | E | RSU | ✓ Infrastructure computes global reputation, verifies blocks, and inserts into a blockchain.<br>✓ CA manages access to the network. | Collects feedback from the regular vehicles. | Reputation is calculated as the average of all ratings. | ✓ Collusion attack. | ✓ Some performance tests.<br>✓ Reputation calculation.<br>✓ Update into the blockchain. | No baseline comparison. | No scenario. | Ethereum test blockchain Ganache |
| [39] | E | RSU | ✓ RSU uses multicriteria decisions to assess reputation.<br>✓ It packs the list of reputations into a block to insert into a blockchain. | RSU collects the trust of the event sender. | RSU calculates the reputation value of the message sender from the trust values of validator vehicles. | No adversary model. | ✓ False message detection rate, malicious vehicle detection latency.<br>✓ Number of dropped false messages, and average trust value. | The false message detection rate is compared with a baseline. | Manhattan grid | OMNeT++ |
| [36] | E | RSU | ✓ Trust-based clustering scheme.<br>✓ RSU selects cluster head.<br>✓ RSU blocks access to malicious vehicles. | Trust propagation. | ✓ RSU calculates the trust from knowledge, experience, and reputation.<br>✓ RSU takes the average of old and new trusts to assign the final trust. | ✓ Wormhole attack.<br>✓ Sybil attack. | Analysis of:<br>✓ Average cluster duration.<br>✓ Average cluster head lifetime.<br>✓ Control overhead, and throughput. | Compared with the two baselines. | Map of Islamabad. | OMNeT++, SUMO |
| [57] | E | RSU | ✓ After the data transmission, a vehicle sends updated trust to the RSU to verify and update into a blockchain.<br>✓ Active detection based malicious vehicle detection. | Probe message, and updated trust from neighbours. | Trust is calculated from the detection trust, reference trust, and transmission trust. | ✓ Packet drop.<br>✓ Spoofing.<br>✓ Cooperation attack. | Analyse of:<br>✓ Delivery ratio.<br>✓ Detection ratio.<br>✓ Average trust value under different ratios of malicious vehicles. | Compared with two baselines in terms of delivery ratio, detection ratio, and average trust value. | Vehicles and pedestrians are randomly distributed in an area of (5000m X 5000m). | ns-3 |
| [58] | E | RSU | ✓ A horizontal trust scheme detects malicious vehicles using different machine-learning algorithms.<br>✓ The vertical trust scheme verifies trust.<br>✓ A blockchain contains the vehicular trust list. | Many votes/decisions are collected about a vehicle. | RSU sums all the votes for a vehicle and checks if it is greater than a threshold to put it into a trusted list. | ✓ Sybil attack. | Analyse of:<br>✓ Accuracy versus amount of collected data using SVM, Random Forest, and KNN. | No baseline comparison. | Traffic scenarios are not mentioned | Ethereum environment. |
| [59] | E | RSU | ✓ A Dirichlet distribution-based trust model and blockchain stores the trust of vehicles.<br>✓ Distributed k-anonymity-based cloaking region maintains the privacy of vehicles. | Recorded historical trust information is queried to update the trust. | Historical trust information of vehicles is added with the trust degree as a requestor and cooperator. | ✓ Bad mouth attack.<br>✓ On-off attack. | ✓ Theoretical security analysis against some attacks.<br>✓ An analysis of malicious vehicle detection in the presence of on-off attack. | The probability of location data leakage and the percentage of maliciousness in the cloaking region are compared with two schemes. | Real driving data is collected for 24 hours from Cologne, Germany. | HyperLeader, JAVA |

| [62] | E | RSU | ✓ The trust model uses Dirichlet regression and punishment mechanism. ✓ Blockchain stores the rating of service providers. | Service rating of a service provider. | ✓ Trust evaluation considers positive, neutral, and negative ratings. ✓ Malicious reviewers and servers are blocked when threshold conditions are met. | ✓ Simple attack. ✓ Slander attack. ✓ Strategic attack. | Analysis of: ✓ Normal/malicious vehicle reputation. ✓ An analysis of collaborative vehicles and success rates versus epoch. ✓ Finds the false positive and false negative. | This model is compared with the beta distribution-based trust model. | 24-hour taxi GPS data collected from Chongqing. | Not mentioned. |

## Table 3: Comparison of Entity Trust Models (Part 2)

| Ref. | Type | Any Road-side units? | Underlying Principle | Feedback Collection | Trust computation | Adversary Model | Analysis | Baseline Comparison | Traffic Scenarios | Simulators |
|---|---|---|---|---|---|---|---|---|---|---|
| [43] | E | RSU | ✓ Bayesian inference based local trust calculation of vehicles. ✓ TrustRank-based algorithm to calculate the global trust of vehicles. ✓ Trust is propagated in a Markov process manner through seed vehicles to other vehicles. | Local trust. | ✓ Local trust is calculated using Bayesian distribution. ✓ Global trust is computed iteratively using the next decaying factor. | ✓ Newcomer attack. ✓ Collusion attack ✓ Bad-mouth attack. ✓ On-off attack. | ✓ Trust evolution in the presence of different rates of maliciousness. ✓ Analysis of true positive rate and true negative rate. | Compared with baselines. | Motorway in Beijing. | Veins, SUMO, OMNeT++ |
| [76] | E | RSU and base stations. | ✓ Trust and reputation approach which isolates malicious users from the network. ✓ Each vehicle computes the trust of other vehicles from which it receives a message. | Feedback | ✓ Trust is computed using the reputation, and recommendations from both neighbouring vehicles and RSU. ✓ Trust about the received message is determined using three fuzzy sets. | ✓ False message. ✓ Collusion attack. | ✓ Accuracy and scalability evaluation with and/or without collusion attack. | Not compared. | Fixed traffic mobility in an area of (100mX 100m). | Bespoke simulator, TRMSIM-V2V |
| [87] | E | Fixed access point. | ✓ Secure authentication based on direct trust and indirect trust calculation scheme. ✓ The correlation coefficient filters out malicious recommendations. | Application data. | ✓ Trust is calculated from the direct and indirect trust. ✓ Indirect trust is calculated from the feedbacks. | ✓ On-off attack. ✓ Bad mouth attack. | ✓ Security degree analysis. ✓ Indirect trust evolution. | No baseline Comparison. | The network scenario is not stated. | MATLAB |
| [54] | D | No RSU | ✓ Concept of the separate event area, decision area and distribution area. ✓ Situation recognition. ✓ Opinion piggybacking. | Indirect reputation. | ✓ Final trust is calculated from the direct and indirect reputation. ✓ Confidence decision is taken based on different situations and reputation constraints. | ✓ Modification attacks. | ✓ Not mentioned. | Not mentioned. | Not mentioned. | Not mentioned. |
| [42] | E | RSU | ✓ This approach has a trust manager, route manager, and decision manager. | Neighbour information. | ✓ Trust computation from past experiences, neighbour information, vehicle trust, and the packet delivery ratio. | ✓ No adversary model. | ✓ Analysis of packet delivery ratio and delay. | Compared with baseline in terms of packet delivery ratio. | A grid topology is used. | ns-2 |
| [88] | E | No RSU | Analytical Hierarchy Process (AHP) based method which utilizes direct and indirect trust. | Feedback. | The previous reputation is added with the direct and indirect trust to get the final trust. | No adversary model. | Analysis of: ✓ Communication delay in the presence of different numbers of vehicles. | No baseline comparison. | A road with multiple unidirectional lanes. | Not mentioned. |
| [89] | E | No RSU | ✓ It finds the trusted routing path using the link quality and trust. ✓ It filters out malicious data. | Not required. | It calculates neighbour trust using role-based trust, recommendation | ✓ DDoS attack. ✓ Selfish behaviour. | Analyse the inserted traffic from various vehicles to detect DDoS attacks and selfish behaviour. | Compared with two other schemes. | 10 Km highway, with two lanes in each direction. | ns-2, VanetMobiSim |

| Ref. | Type | Use of Road-side Units? | Underlying Principle | Feedback Collection | Trust computation | Adversary Model | Analysis | Baseline Comparison | Traffic Scenarios | Simulators |
|---|---|---|---|---|---|---|---|---|---|---|
| | | | | | trust and historical trust. | | | | | |
| [99] | E | Fog network, RSU | ✓ Task-based Experience Reputation (TER). ✓ Fog nodes collaborate to collect and send aggregated trust to RSU. ✓ Differences between frequent versus occasional visitors. | Not required. | Accumulate reputation based on task basis reward and punishment. | No adversary model. | Analysis of: ✓ Communication overhead in the presence of a well-known and experience-based trust model. ✓ Workload of both experience and task-based model. | Compared with experience-based trust. | Random distribution of vehicles. | MATLAB |
| [97] | E | RSU | ✓ A V2V authentication and trust evaluation scheme. ✓ ATA evaluates the trust value of vehicles based on reward points. | Safety Message verification and acknowledgement messages. | Trust is updated by adding the current trust with the reward or punishment for the safety message. | Theoretical robustness against: ✓ Impersonation. ✓ Repudiation. ✓ Message tampering. ✓ Identity disclosure attacks. | Computation and communication overhead under different traffic densities. | Communication overhead is compared with baselines. | A two-lane two-way highway. | ns-3, SUMO, MOVE |
| [100] | E | Fog nodes | ✓ The fuzzy logic-based trust model considers the message lifetime, previous interaction with the sender, and fog node opinion about the event. ✓ Relays a message if the sender is trusted. | Event confirmation from the fog node. | Finds the trust of the sender vehicle from the fuzzy logic approach. | ✓ False message. ✓ Message alteration attack. | Accuracy evaluation. | No baseline comparison. | An area of 2km X 2km. | ns-2, SUMO, MOVE |

## Table 4: Comparison of Data Trust Models (Part 1)

| Ref. | Type | Use of Road-side Units? | Underlying Principle | Feedback Collection | Trust computation | Adversary Model | Analysis | Baseline Comparison | Traffic Scenarios | Simulators |
|---|---|---|---|---|---|---|---|---|---|---|
| [60] | D | RSU | ✓ Both vehicles and RSU need a threshold number of alerts to verify an event. ✓ RSU inserts the validated events into a blockchain. | Collect traffic information to verify an event. | The trust of an event is validated through a threshold-based validation process. | ✓ False attacks. | ✓ Impact on percentage of attackers on the false event success rate. ✓ Analysis of synchronization time requirement of consensus algorithms. | No baseline is mentioned. | Real traffic data collected from vehicle detectors on Taiwan highways. | ns-3 |
| [61] | D | RSU and 5G base station. | ✓ A vehicle uploads a traffic event video on the server with an attached road situation tag which is scored by other vehicles. ✓ RSU authenticates and calculates the trust of sender vehicles. | Tag scores (feedback). | RSU calculates the trust of a tag using the distance between the sender and the scoring vehicles. | ✓ Fake attacks. | Accuracy of malicious vehicle detection, encrypted traffic video overhead, and transmission delay versus message rate analysis. | Not compared. | (1000m X 1000m) area, where vehicles move in a random direction. | OMNeT++, crypto++ |
| [71] | D | RSU | ✓ ML approach uses a pair of Basic Safety Messages (BSMs) to detect location spoofing attacks. ✓ KNN, Random Forest, Decision tree, and Naïve Bayes algorithm are used to detect fake positions. | Not required. | ✓ BSMs from vehicles are analyzed using different machine-learning to classify the source as legitimate or malicious. ✓ Consideration of binary and multiple classifiers. | Location spoofing attack. | Analysis of: ✓ Precision and recall. ✓ F1-score. | Compared with a baseline. | Not considered. | The VeReMi dataset, simulator is not mentioned. |

| Ref. | Type | Use of RSU | Underlying Principle | Feedback Collection | Trust computation | Adversary Model | Analysis | Baseline Comparison | Traffic Scenarios | Simulators |
|---|---|---|---|---|---|---|---|---|---|---|
| [75] | D | RSU | ✓ It encompasses data formalization, trust evaluation and strategic module. ✓ A reinforcement learning model is used to fine-tune the evaluation strategy. | Neighbour information. | An event is verified using both external and internal data. | ✓ Faulty ✓ False message. | Analysis of precision ratio using different numbers of rounds. | Compared with baselines. The proposed scheme is better when the malicious rate is > 50%. | Map of Huangpu district. | OMNeT+, SUMO, Veins |
| [78] | D | RSU | ✓ Tanimoto coefficient is used to find the similarity between the event and beacon. ✓ RSU can check which messages are more trustworthy than others and disseminate its opinion to neighbours. ✓ It is a cryptographic and pseudo-identity-based trust scheme. ✓ RSU calculates the confidence of opinion using a distance-based method. | Opinion. | Trust is calculated from the direct and indirect trust. | ✓ Alteration attacks. ✓ Bogus message attacks. | ✓ F-measure on the threshold. ✓ Malicious vehicle rate and amount of vehicle analysis. ✓ Illustration of detection delay versus malicious vehicle rate and amount of vehicles. | Compared with a baseline to show better decision delay. | Random trips on some streets map. | ns-2 |
| [79] | D | No RSU | ✓ The Tanimoto coefficient is used to crosscheck the beacon message with an alert message to find the higher trustworthiness. ✓ Messages are encrypted and a pseudo-identity is attached while transmitting. | Indirect trust (opinion). | Dempster-Shafer Theorem (DST) is used to combine multiple opinions from neighbours. | ✓ Alteration attacks. ✓ Bogus attacks. ✓ Message suppression attacks. | ✓ Analysis of F-measure under alteration and bogus attacks. ✓ Analysis of location privacy schemes. ✓ Detection delay under alteration attack. | Compared with multiple trust schemes. | Manhattan grid. | ns-2 |

### Table 5: Comparison of Data Trust Models (Part 2)

| Ref. | Type | Use of Road-side Units? | Underlying Principle | Feedback Collection | Trust computation | Adversary Model | Analysis | Baseline Comparison | Traffic Scenarios | Simulators |
|---|---|---|---|---|---|---|---|---|---|---|
| [45] | D | No RSU | ✓ It verifies the location and time closeness. ✓ It has a confidence module, a trust management module, and a decision module. | Not required. | ✓ A receiver vehicle determines the confidence in each unique message and trust of each message about an event. ✓ Then the message with the maximum trust is selected from the decision module. | False information about location and time. | ✓ Effect of malicious nodes on trust and confidence value. ✓ Fake location detection analysis. ✓ Theoretical proofs against malicious behaviour. ✓ False positive rate under various malicious rates. | Not compared with a baseline. | Suffolk county road map. | SWAN++, ONE simulator |

| Ref. | Type | Use of Road-side Unit? | Underlying Principle | Feedback Collection | Trust computation | Adversary Model | Analysis | Baseline Comparison | Traffic Scenarios | Simulators |
|---|---|---|---|---|---|---|---|---|---|---|
| [86] | D | RSU | It employs ant colony optimization which uses both direct observation and feedback to evaluate the trustworthiness of data. | Feedback. | Trust is calculated from the correct data, faulty data, and cooperatively falsified data. | ✓ Cooperatively falsified data attack. | ✓ Analysis of trust under various observing conditions. ✓ Resilience to cooperatively-falsified data attack. ✓ Data delivery delay versus distance and densities. | Not compared with a baseline. | Highway. | ns-3 |
| [93] | D | No RSU | ✓ Weighted voting-based trust model. ✓ The vehicle that is closer to the event has a higher weight. | The vehicle receives opinions only from the in-front vehicles. | ✓ A vehicle decides about an event using a sum of weighted opinions from in-front neighbours. | ✓ Selfish behaviour. | The percentage of incorrect messages is compared for different voting schemes. | Different voting schemes are examined. | Set of road intersections. | NCTUns, C++ |
| [46] | D | No RSU | ✓ Vehicles ignore messages coming from behind. ✓ It considers warning (running state) and traffic events, ✓ It assumes an event and an opposite event. ✓ Forwards a message when it comes from in front and influential area and does not exceed a threshold time. | Not required. | ✓ It verifies the message location and message generated from in front area of a receiver, and it checks whether a vehicle drives through this region later. ✓ The scheme decides from the received message's latest time. | ✓ False message attacks. ✓ Collusion attacks. ✓ Bad mouth attacks. ✓ On-off attacks. | Analysis of trip completion time and number of cheated nodes under different malicious vehicle rates and $CO_2$ emission. | Compared the result with different ML approaches. | Map of Nantong city. | OMNeT++, SUMO, Veins |
| [104] | D | No RSU | ✓ An application-oriented scheme. ✓ It checks whether the message is trusted and then assigns a trust value to each neighbourhood road segment. ✓ Each message is transferred through the safest path towards the destination. | Not required. | The path trust is calculated from multiple dimensions, for example, delay, reliability, security, privacy, and anonymity. | Theoretical proof against: ✓ False message. ✓ Message alteration. ✓ Relaying to another path. | Analysis of: ✓ Packet delivery ratio versus delay ✓ Delay versus speed analysis. | Compared with the two baselines. | Highway. | MATLAB |

## Table 6: Comparison of Hybrid Trust Models (part 1)

| Ref. | Type | Use of Road-side Unit? | Underlying Principle | Feedback Collection | Trust computation | Adversary Model | Analysis | Baseline Comparison | Traffic Scenarios | Simulators |
|---|---|---|---|---|---|---|---|---|---|---|
| [1] | H | RSU | ✓ Sender-side trust evaluation using a TPD inside of vehicle. ✓ TPD rewards/punishes using message accuracy, distance, and time as metrics. ✓ RSU resolves disputes using trusted feedback from clarifiers. | Only when dispute arises (both event and opposite event exist) | ✓ A rule based reward/punishment assessment using message accuracy, distance, and time as metrics. ✓ Fixed RSU Judgement. | ✓ Untrue attacks. ✓ Inconsistent attacks. ✓ Cooperation attacks. | ✓ Detection and thwarting of untrue and inconsistent attacks. ✓ Determining the accuracy of the model with varying malicious vehicles rates (0 to 100) and density. | A reputation approach is considered as a baseline to compare the communication overhead and response time. | Urban area considering varying speed ranges from 30 to 84 mph. | Veins |
| | | | ✓ Considers vehicle running by multiple drivers and individual driver trust management.<br>✓ Considers four class of vehicles. | | | | | | | |
| [18] | H | No RSU | ✓ A distributed tier-based message dissemination scheme.<br>✓ Computes the trust of vehicles which disseminate traffic events.<br>✓ Classify trusted and malicious vehicles. | Not used. | ✓ The receiver vehicle checks the authenticity of the message and assigns trust to the sender. | ✓ Fake message.<br>✓ Message alteration attack. | ✓ Analysis of expected state and probability of staying at a state. | No baseline. | Urban area (4000m X4000m). | Not mentioned. |
| [40] | H | RSU | ✓ Blockchain technology-based Trusted Execution Environment (TEE) is presented.<br>✓ The lower layer of the hierarchy validates messages and blocks whereas the upper layer manages trust, gives incentives as well as performs consensus. | Trust credits are made public. | ✓ The receiver verifies messages with a signed reputation.<br>✓ Each node receives a global trust credit based on the quality of the messages sent and rating results. | Theoretical proof against:<br>✓ Sybil attacks.<br>✓ Message spoofing attacks.<br>✓ Bad mouth attacks. | ✓ Average throughput and response time on message evaluation using TEE and without TEE. | The average latency in their consensus mechanism is compared with two other schemes. | No traffic scenarios are mentioned. | A high-level Performance Evaluation Process Algebra (PEPA) |
| [73] | H | RSU | Deep learning-based driver classification scheme.<br>✓ | Not used. | ✓ The trust in the vehicle is computed from reward points.<br>✓ It classifies fraudulent and non-fraudulent messages/drivers. | No adversary. | ✓ Normalised reward points.<br>✓ Computational overhead versus vehicle density. | Compared with baseline schemes. | Not mentioned. | Python, ns-3, and TensorFlow |
| [77] | H | No RSU | ✓ It first evaluates the received information for correctness using its observations and the sender's trust.<br>✓ Finally, it classifies vehicles and exchanges a list of honest and malicious vehicles. | List of honest and malicious vehicles. | ✓ It finds the trust of the sending node using logistic regression.<br>✓ Trust is updated using the received message from the sender. | ✓ On-off attack. | ✓ Trust evolution.<br>✓ Accuracy of classifying vehicles based on sending events. | Compared with majority voting and weighted voting schemes. | Circular highway. | OMNeT++ |
| [84] | H | No RSU | It has a separate entity and data-centric trust model to achieve secure routing and improve data delivery rate respectively. | Feedback. | ✓ The entity trust is calculated from the direct and recommendation trust.<br>✓ The trustworthiness of data is calculated using utility theory from different factors. | ✓ Blackhole attacks.<br>✓ Selective forwarding attacks. | ✓ Packet delivery ratio.<br>✓ End-to-end delay<br>✓ Path lengths are evaluated. | Compared with different routing protocols. | Km² area where vehicles move on some selected roads. | VanetMobi-Sim |
| [85] | H | No RSU | ✓ Penalty and time-decaying factors are used to understand better relations.<br>✓ Confidence of direct trust is calculated first. | Feedback. | ✓ Node's trust is computed from direct and recommendation trust. | ✓ Not stated. | Analysis of:<br>✓ Direct trust with successful and failed interaction.<br>✓ Packet delivery ratio versus malicious node rate. | Compared with two baseline schemes. | Manhattan mobile model using 5 X 5 grid. | Not mentioned. |
| [94] | H | No RSU | ✓ Adaptive trust and privacy framework.<br>✓ Identity-based cryptographic scheme.<br>✓ Includes both subjective and objective evaluation of the event and updates the sender's reputation. | Opinion | ✓ Weighted voting-based calculation is used for subjective trust.<br>✓ Use transitive trust for entity trust.<br>✓ Upon validation and recognition, the node updates the reputation of another peer. | ✓ Brute force attacks.<br>✓ Fraudulent message. | Detection rate and correct decision rate versus variable number of attackers. | Compared with one baseline. | Streets on 5Km X 5Km area. | ONE |
| [66] | H | No RSU | ✓ For self-organized VANETs.<br>✓ Can revoke malicious nodes and discard fake messages.<br>✓ Fake source, event time detection.<br>✓ Different credit methods are used for urban and highways. | Not used. | ✓ The receiver evaluates the sender's message to update the sender's trust using sender/event location, event time, history of interactions, urban/rural mode, and received event.<br>✓ Select the message with the highest trust when the trust of each unique message is computed. | ✓ False message. | ✓ Theoretical resilience against false location, event and time spreading.<br>✓ Analysis of travel time, $CO_2$ emission, communication overhead, and accuracy by varying numbers of malicious nodes. | Urban versus rural scenarios are compared. | Highway and urban (map of Jeddah). | MATLAB, Veins |

| Ref. | Type | Use of Road-side Units? | Underlying Principle | Feedback Collection | Trust computation | Adversary Model | Analysis | Baseline Comparison | Traffic Scenarios | Simulators |
|---|---|---|---|---|---|---|---|---|---|---|
| [37] | H | RSU | ✓ It uses entity trust in the data-oriented trust evaluation.<br>✓ Considers distinct messages from different senders about an event.<br>✓ Malicious vehicle revocation.<br>✓ The vehicle collects a trust certificate from the Certification Authority (CA) which is included in the messages and extracted as a weight for data trust calculation. | Trust feedback reporting. | ✓ Vehicles visiting the event location confirm or deny the event.<br>✓ The receiver first verifies the message using selected criteria.<br>✓ Then finds the trust of data using a computational method. | ✓ Fake message.<br>✓ Message tampering attack.<br>✓ Trust manipulation attack.<br>✓ Unfair trust feedback. | ✓ The average trust of honest and malicious vehicles.<br>✓ Correct decision of percentage of vehicles.<br>✓ Number of real and fake broadcasts for an emergency event. | Compared with baseline. | Guangzhou Highway. | SUMO |

## Table 7: Comparison of Hybrid Trust Models (Part 2)

| Ref. | Type | Use of Road-side Units? | Underlying Principle | Feedback Collection | Trust computation | Adversary Model | Analysis | Baseline Comparison | Traffic Scenarios | Simulators |
|---|---|---|---|---|---|---|---|---|---|---|
| [90] | H | RSU | ✓ Likelihood function determines the successful or failure interaction.<br>✓ Beta distribution based direct trust.<br>✓ Historical record is saved in the trust certificate. | Trust feedback | ✓ Final trust is computed from weighted sum of direct trust and historical experience record. | ✓ False message.<br>✓ Blackhole attacks.<br>✓ Feedback attacks. | ✓ Trust changes under various malicious rate.<br>✓ Correct decision rate in urban, suburban, and highway. | Baseline comparison on resisting malicious attacks. | Urban, suburban, and highway scenarios with different speed. | Veins |
| [68] | H | RSU | ✓ It builds context about an event cognitively to trust an event.<br>✓ Time, speed, and distance-based outlier detection method. | Opinion. | Trust evaluation module uses experience, role, opinion, and thread-based trust. | ✓ False message. | ✓ Accuracy, error rate, precision, recall, and F1-score analysis.<br>✓ Malicious node detection. | Trust levels are compared with other schemes. | Both urban and rural. | MATLAB |
| [102] | H | RSU | Email and social network-based trust. | Accumulate sender's trust and two-hop trust propagation is allowed. | ✓ If the receiver finds the sender in the trusted list, then the message is trusted, otherwise, the sender's trust is asked from 2 hops distant neighbours.<br>✓ The receiver calculates trust using social interactions and intermediate trust. | No adversary model. | No analysis is conducted. | No baseline comparison. | No scenario is considered. | Not simulated. |
| [91] | H | RSU | ✓ It is based on direct, indirect, event and RSU-based trust.<br>✓ Revokes dishonest nodes collaboratively.<br>✓ The notion of inter-vehicle trust | Positive or negative feedback. | ✓ Each vehicle sends neighbour evaluation to RSU which is used to compute RSU to vehicle trust. | ✓ False message.<br>✓ DoS attacks.<br>✓ Platooning attacks.<br>✓ Message-dropping attacks. | ✓ Average end-to-end delay versus number of nodes.<br>✓ Packet delivery ratio, throughput. | Compared with baseline schemes. | Valencia city map. | ns-2 and SUMO |
| | | | and RSU to vehicle trust. ✓ Collects and combines indirect recommendations about a vehicle. ✓ It evaluates message quality and event effectiveness. | | ✓ Final trust is calculated from vehicle to vehicle and RSU to vehicle trust. | | ✓ Dishonest vehicle detection ratio. | | | |
| [69] | H | RSU | ✓ Gaussian Naïve Bayesian classifier based direct trust and active detection based indirect trust. ✓ Uses cryptography for authentication and integrity. ✓ Blockchain stores certificate and indirect trust. ✓ Evaluates the event using responses from neighbours. | Active detection needs forwarding of two messages to RSU to detect maliciousness. Also, event evaluation requires recommendations from nearby vehicles | ✓ Takes an average of direct and indirect trust to get the final trust. | ✓ Replay attacks. ✓ Collusion attacks. | ✓ Indirect trust computation using active detection. ✓ Vehicle trust computation using Bayesian classifier. ✓ Indirect trust computation by changing number of broadcasts. ✓ Recognition rate of malicious vehicles versus false message sending probability. | Malicious vehicle recognition comparison with existing trust models. | 1 square kilometre area, with vehicle speed 10-12 m/s | Python |

## COMPARISON OF SENDER-SIDE TRUST EVALUATION VERSUS RECEIVER-SIDE TRUST EVALUATION

It is common in current trust models that any driver can announce events in VANETs as the trust of driver is not evaluated at the time of announcements. In existing trust models, a sender vehicle cannot be deemed a trustworthy source until its trust is verified at the recipient vehicles. If receiver vehicles find the message is reliable, they accept it and treat the sender as trusted. On the other hand, receiver vehicles discard messages from an untrusted sender when the evaluation finds the source is unreliable. The evaluation process consumes a lot of resources to decide on an event's validity. These communications from unreliable sources generate traffic surges in the VANET. When an approach requires indirect recommendations for trust evaluation, vehicles may receive false recommendations. The resultant trust validation may be contaminated when an indirect source sends a deceptive advice. Indirect advice are also a concern if forwarded via an untrusted vehicle. To detect fake advice, some models use further filtering schemes, but this increases the level of complexity. Moreover, when vehicles require a higher decision time this can result in vehicles arriving at a hazardous zone which results in more traffic chaos than the original reported incident. For instance, severe traffic congestion can result around an accident if the trust model decision time latency is high, even if the accident incident is announced in a timely manner. In contrast, sender-side evaluation schemes do not need to verify the event message after its arrival. Hence, it offers lower driver decision time compared to the receiver-side methods. Upon reception, receiver vehicles can detour immediately rather than wait for trust metrics to be evaluated, lowering the chaos on roads.

There may be some cases when a vehicle is authorized, but sufficient trust is not yet established. A fake message from a vehicle in this situation could abuse network resources. Furthermore, some models either gather trust information from the locality or globally. However, they do not cope well with rapid topological differences. Receiver-side models are vulnerable to attacks, performance, and complexity issues. Also, some models do not use high-priority messages from official vehicles. Consider a scenario when a recipient vehicle obtains an accident message from another source vehicle. It needs to decide promptly on which road it should drive. For example, it may detour to another route to avoid the hazardous zone without delaying for the trust verification. Or it can begin the trust validation of the sender vehicle. After the assessment, the driver may then find a route to drive next. He/she may use the planned route, or can change the route, as necessary. If the driver acts without awaiting the trust assessment, for a fake message, he/she helps the malicious driver to reach his/her goal. With most existing models, vehicles opt for the second option which demands more time to decide on the event. As the decision is not ready, some vehicles may enter the hazardous zone. This diminishes the impact of an emergency event broadcasting. This delay is a performance problem. The slow decision time can aggravate a situation.

Conversely, in an environment where only trustworthy drivers can announce an event, receiver vehicles do not need to delay for trust verification. Using sender-side trust score evaluation, receiver vehicles are no longer needed to wait for any verdict from other trust entities and a driver can promptly decide about his/her next action. The difference between sender-side evaluation and receiver-side evaluation is depicted in Fig. 2 and Fig. 3. Fig. 2 illustrates the process followed by a receiver-side trust model. Fig. 3 shows the process followed by a sender-side trust model [1]. Both diagrams consider three primary actors: an RSU, the sender vehicle and a receiver vehicle.

In Fig. 2, when a receiver obtains an accident message from the sender it starts its trust evaluation procedure which further requires communication with neighbouring vehicles and RSUs and then employs the collected information in the trust evaluation. If the receiver vehicle finds the data is reliable then it forwards the message again and/or decides whether to change route. Alternatively, if the message is unreliable then the receiver may notify the RSU of this malicious action from the source vehicle to isolate this entity from the network. Conversely, in Fig. 3 this sort of communication is avoided [1]. Whenever a message arrives, the driver of the vehicle can instantly decide whether he should detour or not as he/she is not involved in trust computation when a message arrives. The receiver vehicle also relays the message into the environment to reach others. However, if any receiver visits the event location and detects the event is a false message it may send an untrue attack report to the RSU. The RSU then rules on the validity of the event using feedback from trusted vehicles. The main distinction point is that with sender-side evaluation, receivers do not need further computation and communication which are required for receiver-side based evaluation schemes. This results in reduced communication overhead and lower decision latency for sender-side approaches.

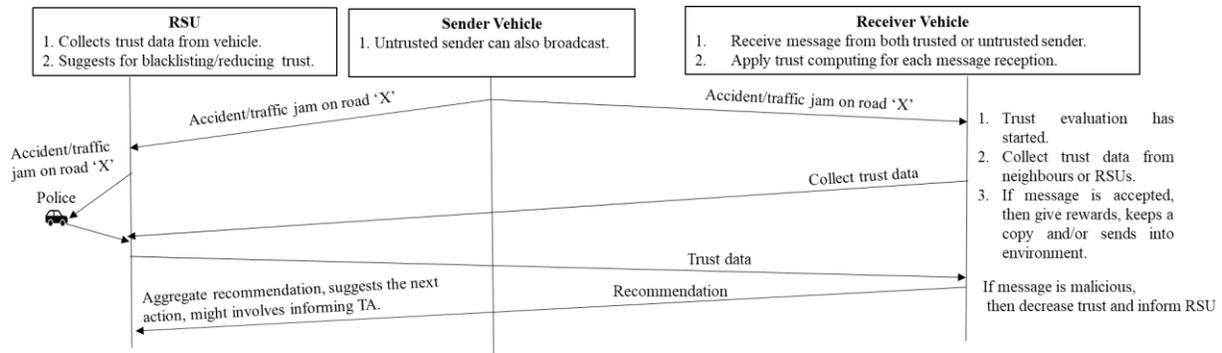

**Fig 2: Sequence Diagram of Event Broadcasting with a Receiver-Side Evaluation-Based Trust Model**

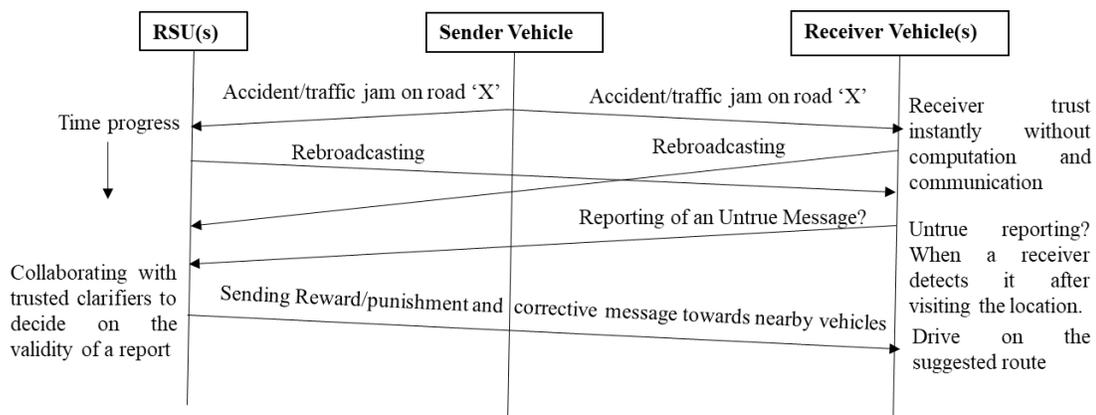

**Fig 3: Sequence Diagram of Event Broadcasting with a Sender-Side Evaluation-Based Trust Model**

## CHALLENGES TO DESIGNING AN EFFICIENT TRUST MODEL FOR VANETS

In VANETs, vehicles use planned routes to reach their destination. However, in the presence of an event, for instance, an accident or congestion along their route, they need to detour, otherwise, they face traffic congestion. If this information is not disseminated and not acted upon on in time, the severity of the conditions may intensify as more vehicles will form a queue around an event. Thus, a VANET should make provision for prompt decision–making. As verification of an event introduces delay, drivers from the surrounding area may enter the hazardous zone and aggravate the congestion.

Event announcements should reach other vehicles quickly, so decisions can be made promptly. Sometimes, these event announcements require further data collection and network analysis to reach a decision. Thus, prompt distribution and forwarding of announcements to neighbours is crucial. While deciding on an event, how many messages are generated and how much time is elapsed to decide on an event, are two vital performance criteria. For example, if vehicles/drivers need to delay 30-45 seconds or more to decide on an event, then the vehicle/driver may bypass the entrance/junction to the only possible detouring option. Additionally, high speed vehicles can reach the event zone soon after the dissemination of event messages. This allows very little time to act appropriately to a situation.

Besides trust evaluation, a trust model also disseminates approach-inherent messages into the VANET which contributes to the communication overhead. For example, many approaches collect trust metrics or recommendations from neighbours and RSUs. Also, whenever receivers initiate further communication after a message arrival this adds latency to the decision-making process. Most existing trust models require a considerable decision time and exhibit high communication overhead as they evaluate messages at the receiver side. A trust model which manages trust with low verification time and communication overhead is more desirable. This is why we consider whether a trust model requires any feedback or recommendation data to decide on the event. The researchers in [1] suggest managing trust at the sender-side to relieve the receivers of the burden of trust verification at run time. This offers near-zero decision delay (ignoring the processing time at the receiving vehicles) which is advantageous over receiver-side evaluation-based trust models. As a result, drivers can decide promptly whether they need to act or not.

In literature, some models assume the use of RSUs whereas others do not. Both options have advantages and disadvantages. For example, if a model can manage trust without using any RSU, then the cost on the network is reduced but lacks authority monitoring. On the other hand, the models which rely on RSUs, suffer from higher cost but offer greater oversight. This offers better decision-making when the decision is carried out by the RSU or authority. Additionally, some models are only application-oriented schemes which lack authority monitoring. The Road Transport Authority (RTA) would decide how they need to manage traffic monitoring by using different hierarchies of authority monitoring centres.

## CONCLUSION

In this paper we have reviewed, classified and summarized trust models for VANETs. We also stated some of the known security threats for VANETs. It is found that trust models apply many different types of technology for example, deep learning and blockchain together. Also, others depend on cloud and blockchain to manage trust. Others consider fog nodes beside RSUs for data verification. It is seen that approaches vary in their trust computational method and the nature of the trust evaluation. Metrics which are common with trust management are direct trust, indirect trust, functional trust, trust from network cooperation, relaying and so on. Some models allow trust computation by receiver vehicles only and others allow RSUs to compute the trust and share the updated trust with the vehicles. In some schemes, only RSUs compute the trust based on the feedback from the neighbouring vehicles. These are receiver-side evaluations of trust which require feedback generation and suffer from latency. Thus, approaches which follow strategy introduce high communication overhead and response times. Alternatively, one can manage trust at the sender vehicle and control network access based on a trust score to relieve the burden of trust computation at the receiver vehicles when a message arrives. This reduces communication overhead and response time. Even so, sometimes highly trusted vehicles can announce untrue messages. These can be verified by the authority, for instance, (nearby RSUs) using trust and feedback from trusted neighbours. Then RSUs can assign appropriate reward and punishment to the drivers who are involved in the dispute (when both a traffic event and the reverse event coexist). Scalability mostly relies on RSU deployment and on the availability of DSRC. When RSUs can cover all roads in an area, then a VANET can offer better traffic management and driving comfort to its users.

**Authors' Profiles**

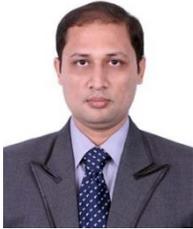

**Rezvi Shahariar** obtained his B.Sc. degree and an M.S. degree in Computer Science and Engineering from the University of Dhaka, Bangladesh in 2006. After some time working as a lecturer at the University of Asia Pacific, Dhaka, Bangladesh, he is now an Associate Professor at the Institute of Information Technology, University of Dhaka. Previously, he worked as an Assistant Professor and Lecturer at the same institute. He received a PhD on trust management for vehicular ad hoc networks at Queen Mary, University of London. His research focuses effective resource management in wireless network environment with an emphasis on trust, security in VANETs, fuzzy modelling, driver behaviour modelling, and the application of machine learning to security

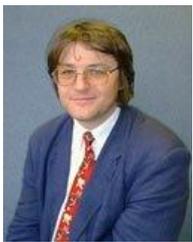

**Chris Phillips** (MIEEE) received a BEng. Degree in Telecoms Engineering from Queen Mary, University of London (QMUL) in 1987 followed by a PhD on concurrent discrete event-driven simulation, also from QMUL. He then worked in industry as a hardware and systems engineer with Bell Northern Research, Siemens Roke Manor Research and Nortel Networks, focusing on broadband network protocols, resource management and resilience. In 2000, he returned to QMUL as a Reader. His research focuses on management mechanisms to enable limited resources to be used effectively in uncertain environments.